\documentclass[fleqn,usenatbib]{mnras}

\usepackage{newtxtext,newtxmath}

\usepackage[T1]{fontenc}

\DeclareRobustCommand{\VAN}[3]{#2}
\let\VANthebibliography\thebibliography
\def\thebibliography{\DeclareRobustCommand{\VAN}[3]{##3}\VANthebibliography}

\usepackage{graphicx}	
\usepackage{amsmath}	

\title[Milky Way satellites, surviving and disrupted]{
Differences in the properties of disrupted and surviving satellites of Milky-Way-mass galaxies in relation to their host accretion histories}
\author[Grimozzi et al.]{Salvador E. Grimozzi$^{1,2}$, Andreea S. Font$^{3}$, Mar\'{\i}a Emilia De Rossi$^{1,2}$\\
$^{1}$Universidad de Buenos Aires, Facultad de Ciencias Exactas y Naturales, Buenos Aires, Argentina\\
$^{2}$CONICET-Universidad de Buenos Aires, Instituto de Astronomía y Física del Espacio (IAFE), Buenos Aires, Argentina\\
\thanks{Corresponding author:A.S.Font@ljmu.ac.uk},
$^{3}$Astrophysics Research Institute, Liverpool John Moores University, 146 Brownlow Hill, Liverpool L3 5RF, UK\\
}

\date{Accepted XXX. Received YYY; in original form ZZZ}

\pubyear{2023}

\begin{document}
\label{firstpage}
\pagerange{\pageref{firstpage}--\pageref{lastpage}}
\maketitle

\begin{abstract}

From the chemo-dynamical properties of tidal debris in the Milky Way, it has been inferred that the dwarf satellites that have been disrupted had different chemical abundances from their present-day counterparts of similar mass that survive today, specifically, they had lower [Fe/H] and higher [Mg/Fe]. Here we use the {\texttt ARTEMIS} simulations to study the relation between the chemical abundances of disrupted progenitors of MW-mass galaxies and their stellar mass, and the evolution of the 
stellar mass -- metallicity relations (MZR) of this population with redshift.  We find that these relations have significant scatter, which correlates with the accretion redshifts ($z_{\rm acc}$) of satellites, and with their cold gas fractions. We investigate the MZRs of dwarf populations accreted at different redshifts and find that they have similar slopes, and also similar with the slope of the MZR of the surviving population ($\approx 0.32$). However, the entire population of disrupted dwarfs displays a steeper MZR, with a slope of $\approx 0.48$, which can be explained by the changes in the mass spectrum of accreted dwarf galaxies with redshift. We find strong relations between the (mass-weighted) $\langle z_{\rm acc} \rangle$ of the disrupted populations and their global chemical abundances ($\langle$[Fe/H]$\rangle$ and $\langle$[Mg/Fe]$\rangle$), which suggests that chemical diagnostics of disrupted dwarfs can be used to infer the types of merger histories of their hosts. For the case of the MW, our simulations predict that the bulk of the disrupted population was accreted at $\langle z_{\rm acc} \rangle \approx 2$, in agreement with other findings. We also find that disrupted satellites form and evolve in denser environments, closer to their hosts, than their present-day counterparts. 
\end{abstract}




\section{Introduction}
\label{sec:intro}

In the standard $\Lambda$CDM cosmological model, large galaxies like the Milky Way (MW) grow hierarchically, by accreting many smaller galaxies over time \citep{White1991}. In MW-like galaxies, signatures of accreted and disrupted dwarf galaxies can be recovered from the properties of the tidal debris found at present-day in the stellar halo \citep{Bullock2005,Cooper2010,Monachesi2019,Helmi2020}. Many observations support this hierarchical paradigm of galaxy formation, numerous tidal streams having being already discovered in the MW (e.g., \citealt{Helmi1999,Newberg2003,Majewski2004,Belokurov2007,Myeong2019,Li2022,Naidu2020,Yuan2020}), in M31 (e.g., \citealt{Martin2014}), and in more distant spiral galaxies (e.g., \citealt{Martinez-Delgado2010}). 

Combining kinematical and chemical abundance data for stars in the tidal debris is deemed to be a powerful tool for pinning down the properties the progenitors, and hence, to reconstruct the merger history of the host galaxy \citep{McWilliam1997,Freeman2002}. Previous numerical simulations have shown the validity of this approach and were able to put some constraints on the merger history of the Milky Way (e.g., \citealt{Font2006b,Johnston2008,Gilbert2009,Lee2015,Cunningham2022,Deason2023}). However, to date, the full reconstruction of MW's merger history is limited by the availability of the observational data. Reconstructing the full merger history will also help us place our Galaxy within the larger cosmological context, and determine whether this history is ``typical'' among galaxies of similar mass \citep[e.g., ][]{DeRossi2009}. 

For the Milky Way there are now exquisite kinematical and chemical abundance data for many stellar tidal streams, thanks to precise astrometric measurements from Gaia \citep{Gaia2018} and high-resolution spectroscopy from the APOGEE \citep{Majewski2017}, GALAH \citep{DeSilva2015} and H3 \citep{Conroy2019} surveys. This wealth of data has enabled the reconstruction of many of the past accretion events. For example, it is now well established that a massive dwarf galaxy, called Gaia Enceladus/Sausage (GES), was accreted by the MW about $8-9$~Gyr ago \citep{Belokurov2018,Helmi2018}. It is estimated that, at the time of accretion, GES had a stellar mass similar to that of the Large Magellanic Cloud (LMC) today. Recently, streams from other, less massive, disrupted satellites have been uncovered by  observations. These include the Kraken \citep{Kruijssen2019}, Sequoia \citep{Myeong2019}, Thamnos \citep{Koppelman2019}, Wukong and I'toi \citep{Naidu2020}. However, although the main branches of the MW's ``merger tree'' have been identified, many other tidal streams may still remain hidden. Cosmological models predict that the disrupted dwarf galaxies make up the vast majority ($\sim 70-80\%)$ of all the satellites ever accreted onto the MW \citep{Fattahi2020,Santistevan2020}. Given that the current census of surviving satellites is $\ga 50$ \citep{Drlica-Wagner2020}, this suggests that the merger history of the MW is still to be determined. 

With a growing number of tidal streams from disrupted satellites, we begin to have a better picture about the properties of these early accreted dwarf galaxies, and how these compare with the properties of present-day satellites. The emerging picture is that disrupted dwarf galaxies (i.e., the early progenitors, or the MW ``building blocks'') had different evolutionary paths than their present-day counterparts. For example, by using data from Gaia and the H3 survey, \citet{Naidu2020} were able to assign a good portion of the stellar halo to individual tidal streams from disrupted galaxies. With these data, \citet{Naidu2022} found that at fixed stellar mass, disrupted satellites have lower metallicities than the surviving ones, ([Fe/H] lower by $\simeq 0.4$~dex). They also found that, at fixed [Fe/H], the disrupted satellites are more $\alpha$-enriched than present-day satellites ([Mg/Fe] ratios higher by $\sim 0.2-0.3$~dex).

These results are consistent with the well known chemical abundance differences between the stars in the halo of the MW and stars in present-day satellites. At fixed [Fe/H], the stellar halo shows higher [$\alpha$/Fe] values than those in surviving dwarf galaxies, typically by $0.2$~dex \citep{Fulbright2002,Shetrone2003,Venn2004,Tolstoy2009}. This suggests that the surviving population is unlike the MW's building blocks and that it has followed different star formation and chemical evolution paths than the disrupted population. Simulations run in $\Lambda$CDM models are able to explain these differences, showing that the early accreted dwarfs tend to have more bursty star formation histories than the later accreted ones \citep{Robertson2005,Khoperskov2023}. Consequently, the accreted component of the MW stellar haloes in these simulations have higher [$\alpha$/Fe] than surviving satellites which, having been accreted later, had more time to form stars and increase their Fe content, and therefore have lower [$\alpha$/Fe] abundances (see, for example, \citealt{Font2006a}).

The observed differences in [Fe/H] between the disrupted and surviving populations prompt the question whether they also display different stellar mass -- metallicity relations (MZR). The lower [Fe/H] at fixed stellar mass displayed by disrupted satellites implies that they obey a MZR with a lower normalisation than corresponding relation for surviving satellites. Determining the offset in normalizations between the MZRs of the two populations, and also, if there are any differences in the slopes of these relations, allows us to understand further the intrinsic differences between the two populations, and even to constrain the accretion history of the MW. 

The MZR of surviving population is well determined. Present-day dwarf galaxies follow the same global scaling relation for other, more massive, galaxies \citep{Grebel2003,Kirby2013,Kirby2020}, while the relation is also observed to continue down to the mass regime of the ultra-faint dwarfs, with the same slope and normalisation as measured for the classical ones \citep{Simon2019}. Moreover, the MZR has been observed for galaxies outside the Local Group \citep[e.g., ][]{Gallazzi2005} and at also at higher redshifts \citep[e.g., ][]{Cullen2019}.
State-of-the-art cosmological simulations obtain that the present-day stellar MZR extends over several decades in galaxy mass (e.g., \citealt{Schaye2015, DeRossi2015,DeRossi2017}).  In contrast, the MZR for the disrupted population of MW satellites was characterized only recently. By reconstructing the properties of disrupted dwarfs from the information encoded in the tidal debris, \citet{Naidu2022} found that their MZR has a similar slope to that of the MZR of surviving population ($\sim 0.3$). The MZR of the disrupted population is offset lower in [Fe/H], by $\approx 0.4$, compared with the MZR at present-day from \citet{Kirby2013}. As previously mentioned, \citet{Naidu2022} also find that disrupted satellites are more enriched in [Mg/Fe] than surviving dwarfs of, at either fixed stellar mass, or fixed [Fe/H]. In addition, they find that data supports a scenario in which the disrupted dwarfs formed closer to the MW host. The higher density environment in which progenitors evolved, coupled with the proximity to the MW, are factors that may have led to the more enhanced star formation histories of these early accreted satellites compared with the surviving population.

In this study, we use the \texttt{ARTEMIS} suite of high-resolution cosmological hydrodynamical simulations to compare the properties of the surviving and disrupted satellites of MW-mass galaxies and test this scenario. In addition, we investigate each population in more detail, for example by analysing the scatter in various properties at $z=0$  (e.g., in the [Fe/H] --$M_*$ and [Mg/Fe] -- $M_*$ relations, in the MZRs, and in the [Mg/Fe] -- [Fe/H] distributions). We also study the dependence on redshift of the MZR of disrupted population and compare the results from simulations with available observations in the MW.  

The paper is organised as follows. Section~\ref{sec:sims} summarizes the main characteristics of the suite of MW-mass simulations used in this study. In Section ~\ref{sec:results}, we analyse the (stellar) MZRs of the surviving and disrupted dwarf populations (\S~\ref{ssec:MZR}) and the corresponding [Mg/Fe] -- ${M}_{*}$ and [Mg/Fe] -- [Fe/H] distributions (\S~\ref{ssec:alpha}); we investigate the nature of the scatter in these relations (\S~\ref{ssec:scatter}); and seek to find which properties of the host progenitors (namely, the average chemical abundances, the slope of the MZR, the slope of the [Mg/Fe] -- ${M}_*$ relation) correlate with the accretion histories of their hosts (\S~\ref{ssec:av_prop}); we compare the stellar mass distributions of surviving and disrupted dwarfs (\S~\ref{ssec:mass_dist}); we also follow the locations in which the disrupted and surviving satellite populations formed and evolved prior to accretion onto their hosts (\S~\ref{ssec:environment}), in order to determine the role that  environment plays in their properties; and analyse the disruption times of the disrupted progenitors after accretion onto their hosts (\S~\ref{ssec:disruption}). Section \ref{sec:discussion} includes a discussion of our results, and Section \ref{sec:concl} presents the conclusions of this study. 

Unless otherwise specified, all chemical abundances correspond to the stellar components of dwarf satellites, while MZR refers to the stellar mass -- {\it stellar} metallicity relation.

\section{\texttt{ARTEMIS} simulations and merger trees}
\label{sec:sims}

\texttt{ARTEMIS} is a suite of zoomed-in, cosmological hydrodynamical simulations of MW-mass systems run in a flat $\Lambda$CDM WMAP9 \citep{hinshaw2013} model. The following cosmological parameters are assumed: $\Omega_\textrm{m}=0.2793$, $\Omega_\textrm{b}=0.0463$, $h=0.70$, $\sigma_8=0.8211$ and $n_s=0.972$. Here, we use the sample of $42$ MW-mass hosts  simulations introduced in \citet{Font2020}, with total masses  ranging between $8 \times 10^{11} < {M}_{200}/{\rm M}_{\sun} < 2 \times 10^{12}$, where ${M}_{200}$ is the mass enclosing a mean density of $200$ times the critical density of the Universe at present time. Dark matter particle masses are $1.17\times10^5$ M$_{\sun}\/h^{-1}$ and the initial baryon particle masses are $2.23\times10^4$ M$_{\sun}\/h^{-1}$, while the force resolution (the Plummer-equivalent softening) is $125$~pc $h^{-1}$. Below we summarise the main properties of the simulations.

\subsection{Subgrid physical prescriptions}

\texttt{ARTEMIS} simulations were carried out with the Gadget-3 code (last described in \citealt{Springel2005}), with an updated hydro solver and subgrid physical prescriptions developed for the \texttt{EAGLE} project \citep{Schaye2015}. The latter include  metal-dependent radiative cooling in the presence of a photo-ionizing UV background, star formation, stellar and chemical evolution, formation of supermassive black holes, stellar feedback from supernova (SN) and stellar winds, as well as feedback from active galactic nuclei. These prescriptions are described in detail in  \cite{Schaye2015} and \cite{Crain2015} (see, also, references therein). 
In particular, the \texttt{ARTEMIS} chemical enrichment model considers the yield from AGB stars, stellar winds and core collapse SN. 11 element species are followed: H, He, C, N, O, Ne, Mg, Si and Fe are tracked individually, and Ca and S track the Si abundance. The yield depends on the amount of mass reaching the end of the main sequence phase in each star particle. Nucleosynthetic yields from \citet{Marigo2001} and \citet{Portinari1998} are applied. The yield from SNIa is implemented separately, where the rate of SN events per unit initial mass is defined as $N_{\rm{SNIa}}=\nu {\tau}^{-1} e^{-t/\tau}$, with $\tau=2~\rm{Gyr}$ and $\nu=2\times 10^{-3}~\rm{M_{\sun}}^{-1}$ (see \citealt{Wiersma2009} and \citealt{Schaye2015}). The SNIa delay time considered in this model takes into account the minimum time that is required for stars to evolve into white dwarfs\footnote{As discussed in \citet{Wiersma2009}, there are other prescriptions that could model the rate of SNIa (see also \citealt{Maoz2010}). We note that changes to the SNIa delay time in the model can lead to variations in chemical properties of galaxies, such as their $\alpha$-enhancement. An analysis of this effect is beyond the scope of this paper.} The SNIa yields are from \citet{Thielemann2003}. 

The stellar feedback scheme used in \texttt{ARTEMIS} is similar to that implemented in the \texttt{EAGLE} simulations \citep{Schaye2015}, however its associated parameters were re-calibrated (see \citealt{Font2020}) to obtain a better match to the stellar mass -- halo mass relation for galaxies in the MW-mass range.  As a result, the host galaxies in the \texttt{ARTEMIS} sample encompass the range of stellar masses and magnitudes of a large number of ``MW analogues'' from the Local Group (i.e., M31), the Local Volume and in the SAGA survey (10 -- 40\, Mpc; \citealt{Geha2017}). For further details, see \citet{Font2021}. 
The global properties of surviving satellites around MW analogues are also matched remarkably well \citep{Font2021,Font2022}. These include: the overall number and the radial distribution around their hosts, their luminosity functions, colours, current star formation rates and quenched fractions. The simulations match these observables in the surviving satellites in the MW, M31 and in other MW analogues (e.g., \citealt{Bennett2019,Carlsten2021a,Carlsten2021b}). 

In the simulations, individual galaxies (MW-mass hosts and dwarf satellites) are identified as gravitationally bound structures (haloes and subhaloes, respectively) using the \texttt{SUBFIND} halo finding algorithm described in \citet{Dolag2009}.

\subsubsection{Chemical abundances offsets}
As discussed in \citet{Font2020}, these simulations produce satellite galaxies that are more metal-rich (in both [Fe/H]) and [Mg/Fe]) than observed ones (see figure $2$ of that study). A similar, albeit slightly higher,  discrepancy in metallicities is seen in the \texttt{EAGLE} simulations (see figure $13$ in \citealt{Schaye2015}, and also the discussion in \citealt{DeRossi2017}). Upon closer inspection in \texttt{ARTEMIS}, we find that the chemical abundance values, both [Fe/H] and [Mg/Fe], are {\it systematically} higher than the observed values across several orders in stellar mass ($M_* \simeq 10^6 - 10^{10}\, {\rm M}_{\sun}$). This results in the MZR of surviving satellites being offset systematically upwards in [Fe/H] by about $0.4$~dex, compared with the observed MZR (\citealt{Kirby2013}) in the MW. Similarly, the averaged [Mg/Fe] values for the simulated dwarfs at $z=0$ are higher by $\simeq 0.2$~dex compared to observations.

The cause of these discrepancies between the observed and simulated chemical abundances is not clear, however, given these are systematic  differences across about the four orders of magnitude in stellar mass investigated here, it suggests that the cause may be related to the particular nucleosynthetic yields adopted in these simulations.  As discussed in detail in
\citet{Wiersma2009}, nucleosynthetic yields are uncertain by factors of a few and the evolution of simulated nuclei abundance is sensitive to the specific choice of yield tables. On the other hand, the stellar masses and metallicities inferred from observations may also be affected by systematic uncertainties associated, for example, with the aperture of instruments, selection effects, or the different methods applied for deriving element abundances from the spectral energy distributions (SEDs) of galaxies (e.g., \citealt{Maiolino2019}). 

Because we are mainly interested in the differences between the chemical abundances of surviving and disrupted dwarf galaxies at any given redshift, and since these abundances are affected in a similar manner by the nucleosynthetic yields, the main results of this study are not affected by these systematic differences. To simplify the comparisons with the observations, we choose to subtract a fixed value of $0.4$~dex from all [Fe/H] values of individual star particles in the simulations, in order to match the $z=0$ MZR of observed satellites in the MW. Similarly, we substract a fixed value of $0.2$~dex from all [Mg/Fe] values in the simulations. Hereafter, all [Fe/H] and [Mg/Fe] values include these offsets.  As the same shifts are applied at all redshifts ($z$), this procedure does not affect the study of chemical abundance variations across cosmic time, or the scatter, which are the main focus of this paper.

\begin{figure}
\includegraphics[width=0.95\columnwidth]{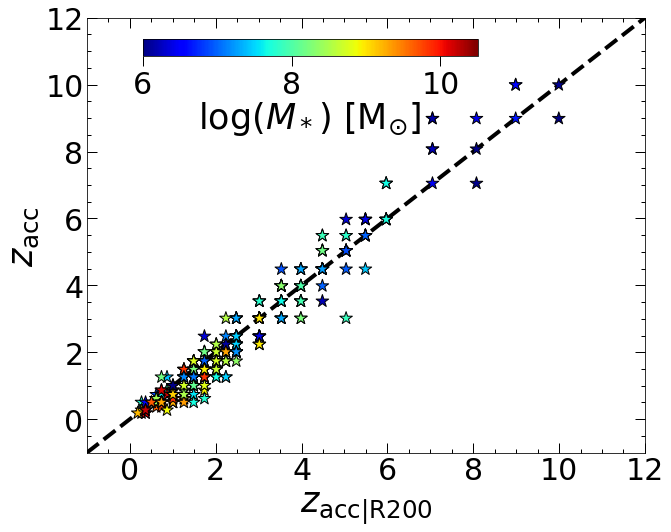}
\caption{The redshift of accretion ($z_{\rm{acc}}$) of disrupted dwarfs versus the redshift when these dwarfs cross the ${R}_{200}$ radius of its MW-mass host ($z_{\rm{acc|{\rm R}_{200}}}$). Values are colour-coded by the maximum ${M}_*$ reached by the dwarfs, i.e., measured at $z_{\rm acc}$. The blue dashed line represents the 1:1 correspondence. In general, the two redshifts are similar for most of the disrupted dwarfs. However, for the majority of the most massive satellites, $z_{\rm{acc}}<z_{\rm{acc}|{\rm R}_{200}} $, which indicates that these continue to form stars for some time while within the ${R}_{200}$ boundary.}
\label{fig:zacc_vszacc200_Mstar}
\end{figure}

\subsection{Tracking dwarf progenitors along merger trees}
\label{sec:merger_trees}

Merger trees for the MW-mass hosts were constructed previously, using the same methods described in \citet{McAlpine2016} for the \texttt{EAGLE} simulations. The $42$ MW-mass systems used here have a variety of merger histories, ranging from quiescent to very active. We use these merger trees to follow the dwarf satellite galaxies back in time, prior to accretion onto their hosts, and to determine their properties over time.
From the merger trees, we identify satellite dwarf galaxies as subhaloes that merge with the main branches (the main haloes, i.e., the main progenitors of the MW-mass galaxies). We define as {\em surviving} all dwarf galaxies that lie at present time ($z=0$) within the virial radius of their host, namely within ${R}_{200}$, which is the radius where the overdensity of the main halo is $200$ times the critical density of the Universe. All other dwarfs that were accreted onto their hosts, but they could not be identified anymore by \texttt{SUBFIND}, are considered to be {\em disrupted}. Note that \texttt{SUBFIND} imposes a minimum number of total particles (dark matter + baryonic), in this case assumed to be 20, in order to identify a self-bound subhalo.

\begin{figure*}
\includegraphics[width=1.9\columnwidth]{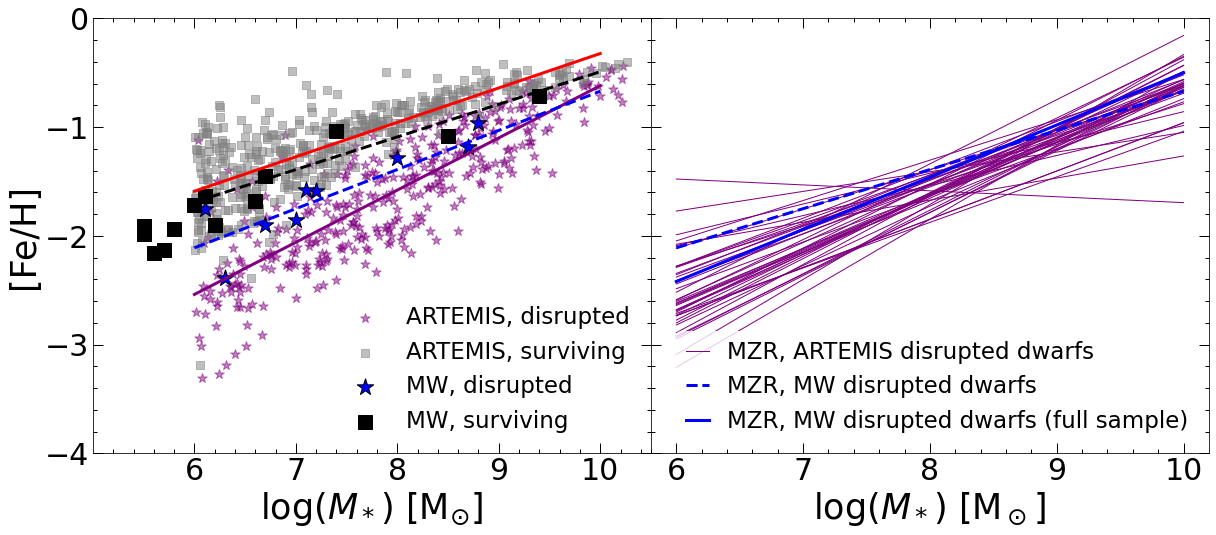}
\caption{{\it Left:} The MZR of disrupted (purple stars) and surviving (grey squares) populations of dwarf galaxies in all \texttt{ARTEMIS} Milky Way-mass hosts. The linear fits to the distributions are shown with solid purple and red lines, for disrupted and surviving populations, respectively. The dashed black line corresponds to the MZR fit of \citet{Kirby2013} for surviving satellites in the Local Group: ${\rm [Fe/H]}=0.30\,\log(M_*/10^{6}~\rm{M_{\sun}})-1.69$.
 The dashed blue line represents the fit from \citet{Naidu2022} for the disrupted population in the MW: ${\rm[Fe/H]}=0.36\,\log(M_*/10^{6}~\rm{M_{\sun}})-2.11$.
{\it Right:} Individual fits to the disrupted dwarf galaxies in each of the simulated MW hosts (purple lines). The two fits for the disrupted population in the MW, from \citet{Naidu2022}, are indicated with blue lines: the dashed blue line (which is identical with the one in the left panel) is derived from a sample which excludes I'itoi and Sagittarius satellites (see text), while the solid blue line includes these satellites in the fit.}
\label{fig:MZR-scatter-fits}
\end{figure*}

Throughout the paper, we refer to two specific redshifts, which are relevant in the evolution of dwarf satellites:

\begin{itemize}
\item For disrupted dwarfs, we refer to their {\em redshift of accretion}, $z_{\rm acc}$. Typically, $z_{\rm acc}$ is defined as the redshift when the subhalo joins the FoF group of the main halo progenitor. Here, however, we define $z_{\rm acc}$ as the redshift when the dwarf galaxy (subhalo) reaches its maximum ${M}_*$. In general, the two definitions give similar results, as seen in Fig.~\ref{fig:zacc_vszacc200_Mstar}. The almost 1:1 correspondence in this figure indicates that most dwarfs reach their maximum mass around the time they cross the ${R}_{200}$ radius of their host (we denote $z_{{\rm acc}| {R}_{200}}$ as the corresponding redshift), after which their star formation quickly becomes quenched. Also, many satellites begin to be tidally disrupted after they enter their host system and experience mass loss. Therefore, for most dwarf satellites, the redshift when they reach the maximum $M_*$ is also when they cross the virial radius ($z_{\rm acc} \approx z_{{\rm acc}| {R}_{200}}$). However, for a small fraction of dwarf galaxies, this is not the case, as they continue to form stars after crossing ${R}_{200}$ and then become disrupted. Fig.~\ref{fig:zacc_vszacc200_Mstar} shows that this occurs mostly for high mass satellites (systems for which $z_{\rm acc} < z_{{\rm acc}| {R}_{200}}$). Since in this study we are interested in the {\it intrinsic} differences in chemical abundances between different populations of dwarf galaxies, we aim to capture the entire chemical evolution of these systems. This is more appropriately represented by the redshift when galaxies reach their maximum $M_*$, i.e., $z_{\rm acc}$. 

\item For surviving satellites, we also compute the redshift when the dwarfs reach half of their maximum $M_*$, denoted as $z_{50}$ (specifically, when $M_*$ is within $40$ to $60~\rm{per}~\rm{cent}$ of the maximum $M_*$). Here we use $z_{50}$ instead of $z_{\rm acc}$ because we aim to determine whether the scatter in chemical abundances of satellites depends on the accretion history of the host. But, as surviving dwarfs are predominantly accreted recently (usually, at $z\lesssim 1$), $z_{\rm acc}$ does not have sufficient range to correlate with any scatter in the properties of this population. Typically $z_{50} > z_{\rm acc}$, given that satellites grow most of their $M_*$ prior to accretion onto their host. 
We note that for a small number of surviving satellites ($45$ out of a total of $551$; i.e., $\approx 8~\rm{per}~\rm{cent}$) we could not identify $z_{50}$, due to an insufficient number of simulation outputs to capture the growth between $40-60~\rm{per}~\rm{cent}$ of the maximum $M_*$. These systems are excluded from the plots that use $z_{50}$ values (Figs.~\ref{fig:MZR-z50-fSF} and~\ref{fig:alpha-Mstar-z50-fSF}). 
\end{itemize}

\section{Properties of disrupted and surviving satellites}
\label{sec:results}

In the following, we compare the properties of simulated dwarf galaxies accreted onto the MW-mass systems, both disrupted and surviving at present day, with the corresponding satellite populations in the MW. For observations, we use the data from \citet{Naidu2022}, comprised of 9 disrupted satellites and 13 surviving, with properties collected from a variety of sources and listed in Tables 1 and 2 of their article (see also references therein). Specifically, the observed disrupted dwarf galaxies are Sagittarius, GES, Helmi Streams, Sequoia, Wukong/LMS-1, Cetus, Thamnos, I'itoi and Orphan/Chenab, while the surviving satellites are the Large and the Small Magellanic Clouds (LMC and SMC), Fornax, Leo I, Sculptor, Antlia 2, Leo II, Carina, Sextans, Ursa Minor, Crater 2, Draco and Canes Venatici I. We note, however, that the observational sample does not include all disrupted satellites, given the spatial limitation of the H3 survey (heliocentric distances of $3-50$~kpc). Because of this, progenitors like Kraken, for example, are not included here. For simulations, we choose to include all disrupted progenitors, as we intend to uncover general trends related to their host accretion histories.

\subsection{Stellar mass -- metallicity relations}
\label{ssec:MZR}

We first compare the MZRs of the surviving and disrupted populations of all $42$ simulated MW-mass systems with the corresponding observed MZRs for the MW. The left panel of Fig.~\ref{fig:MZR-scatter-fits} shows the distributions of mean [Fe/H] versus ${M}_{*}$ for all\footnote{A very small fraction disrupted dwarfs (not shown here) lie far off from the main MZR, showing low ${M}_{*} < 10^8~\rm{M_{\sun}}$ and high $\rm{[Fe/H]}\gtrsim-1.5$. Such subhaloes are usually identified in one or two snapshots and have little or no dark matter (dark matter masses are less than $15~\rm{per}~\rm{cent}$ of their stellar masses). We therefore consider these subhaloes to be spuriously identified by \texttt{SUBFIND} and we further exclude them from our sample.} dwarf galaxies in the full simulated sample. For the disrupted populations in the simulations, the ${M}_{*}$ and [Fe/H] values are computed at $z_{\rm acc}$ (i.e., when ${M}_{*}$ reaches its maximum value, according to our definition in Section~\ref{sec:merger_trees}). For the surviving populations the values correspond to $z=0$, as in the observations. In agreement with the observations, simulations predict well-defined correlations between [Fe/H] and ${M}_*$.  This is seen in both the surviving and disrupted populations. Also, at fixed ${M}_*$, the surviving satellites are typically more metal-rich than the disrupted dwarfs, also in qualitative agreement with the observations. 

For a more quantitative comparison with the observations, we compute the best linear fits for the MZRs of the two dwarf populations in the simulations. Considering the entire dwarf populations in the $42$ MW-mass systems, we obtain: 

\begin{equation}
[{\rm{Fe/H}}]=(0.32\pm0.01)\log(M_*)-(3.49\pm0.10),
\label{eq:MZRfit_s}
\end{equation}

\noindent for the surviving population (solid red line in left panel of Fig.~\ref{fig:MZR-scatter-fits}), and

\begin{equation}
   [{\rm{Fe/H}}]=(0.48\pm0.02)\log(M_*)-(5.42\pm0.13), 
\label{eq:MZRfit_d}
\end{equation} 

\noindent for the disrupted one (solid purple line in the same left panel).

Overall, the MZR fit for surviving satellites in simulations agrees with the one derived for the surviving satellites in the MW. For example, the slope of $0.32\, (\pm 0.01)$ in the simulations is similar to the slope corresponding to the observed sample from \citet{Naidu2022}, which is $\approx 0.36$, and it is also very close to the slope of $\approx 0.3$ derived from a larger sample of Local Group satellites by \citet{Kirby2013} (the latter fit is shown with dashed black line in the left panel of Fig.~\ref{fig:MZR-scatter-fits}). 
We recall that, for simulations,  all [Fe/H] values were adjusted downwards by $0.4$~dex, therefore the normalisation of the MZR in simulations matches the one for the observed satellites in the MW by construction. However, no additional recalibration was done to match the slope of the observed MZR.

The MZR fit for the entire disrupted population in simulations has a slope of $0.48 \pm 0.02$, which marginally agrees (within the errors) with the slope of $0.36^{+0.12}_{-0.04}$ reported by \citet{Naidu2022} for the disrupted population in the MW. We note, however, that the latter value is derived from a sample of 7 disrupted dwarfs, which excludes Sagittarius and I'itoi (the \citealt{Naidu2022} fit is shown with dashed blue line in Fig.~\ref{fig:MZR-scatter-fits}). According to these authors, Sagittarius ($M_* =10^{8.8}\, {\rm M}_{\odot}$, [Fe/H]$=-0.96$) is excluded because it is not yet fully disrupted and was still forming stars as recently as $2$~Gyr ago, while I'itoi ($M_* =10^{6.3}\, {\rm M}_{\odot}$, [Fe/H]$=-2.39$) appears to be an outlier in the relation, its metallicity being too low. However, by taking into account these two systems, the MZR for disrupted dwarfs in MW has a slope of $0.50 \pm 0.08$, which is in better agreement with the average value of $0.48$ predicted for this population by our simulations. One determining factor for improving the comparison between simulations and observations is to better constrain the properties of I'itoi. Furthermore, discovering more debris from low-mass progenitors, similar to I'itoi, would help pin down the slope of the MZR for the disrupted population in the MW.  Also, as mentioned previously, the observational sample of disrupted galaxies of \citet{Naidu2022} does not include systems whose debris are contained near the centre of the Galaxy ($\la 6$~kpc) or in the outskirts ($\ga 50$~kpc), while our simulated sample includes all disrupted dwarfs. A more detailed comparison with observations, taking into account the location of merger debris and other observational selection effects, is left for a future study.

We note that eqs.~\ref{eq:MZRfit_s} and ~\ref{eq:MZRfit_d} correspond to average fits for the entire populations of all simulated MW-mass systems, and significant variations can occur on a system-by-system basis. The right panel of Fig.~\ref{fig:MZR-scatter-fits} shows the different MZR fits for all disrupted populations in each of the simulated MW-mass hosts  (purple lines). For comparison, the linear fits for the disrupted population in the MW from \citet{Naidu2022} are shown here, both excluding (dashed blue line) and including (solid blue line) Sagittarius and I'itoi. This figure shows that the two MZR slopes derived for the disrupted population in the MW ($\approx 0.36$ or $\approx 0.50$, respectively) can be matched by several MW-mass systems. However, we can find more matches in the simulated sample of MW-mass systems to the observations, when the latter includes I'itoi and Sagittarius (solid blue line). This also indicates that the MW may have a fairly typical disrupted population for galaxies of its mass. 
 
On the other hand, matching the MZRs of disrupted and surviving populations simultaneously in the simulations reduces the number of matches, as the MZR fits may differ in normalisation or slope than the values inferred from the observations. In addition, as it will be discussed in the next sections, fully realistic simulations should match other observables besides the MZRs, such as the [Mg/Fe] -- [Fe/H] -- $M_*$ distributions of the two populations.
 
\begin{figure}
\centering
\includegraphics[width=0.95\columnwidth]{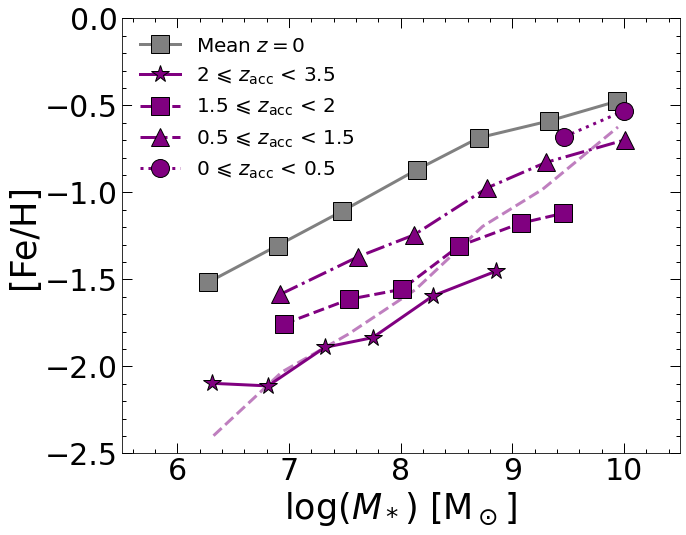}
\caption{Mean MZR for surviving dwarfs at $z=0$ (grey line) and disrupted dwarfs accreted at different redshifts, $z=z_{\rm acc}$ (purple lines). The translucent dashed purple line corresponds to the mean relation for all disrupted dwarfs. [Fe/H] are mean values computed at different $z$ and in different ${M}_*$ bins. For disrupted dwarfs, the normalization of the MZR increases towards lower $z_{\rm acc}$, by about $0.2-0.25$~dex per $z_{\rm acc}$ bin. Although disrupted dwarfs in different $z_{\rm acc}$ bins trace MZRs with similar slopes  ($\approx 0.27 - 0.29$), the slope for the entire disrupted population is significantly steeper ($\approx 0.48$).}
\label{fig:Disrupted-MZR-different-z-single-panel}
\end{figure}

\begin{figure*}
\includegraphics[width=1.9\columnwidth]{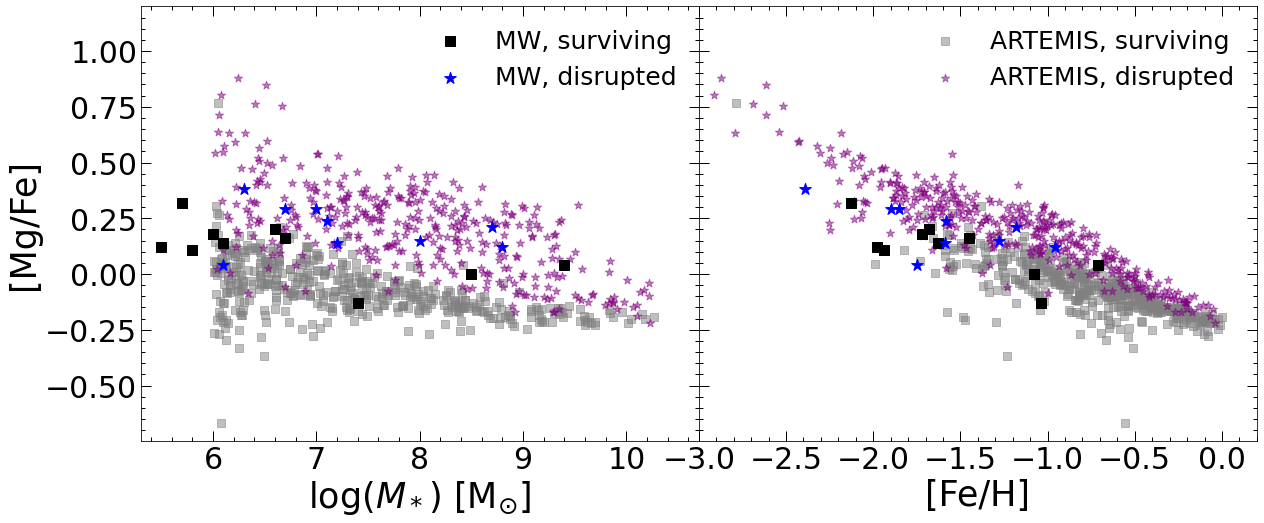}
\caption{{\it Left:}  [Mg/Fe] -- ${M}_*$ distributions for the two populations of dwarfs in simulations, disrupted (with purple stars) and surviving (with grey squares). For comparison, the corresponding populations of the MW from \citet{Naidu2022} are shown, disrupted (with blue stars) and surviving (with black squares). At fixed $M_*$, disrupted dwarfs tend to be more enhanced in [Mg/Fe] than the surviving ones, in both simulations and observations. {\it Right:} A similar comparison for the [Mg/Fe] -- [Fe/H] distribution. At fixed [Fe/H], disrupted dwarfs tend to have higher [Mg/Fe] than surviving ones, again, in both simulations and observations.}
\label{fig:alpha-vs-Mstar}
\end{figure*}

For the disrupted population, we also investigate the evolution of its MZR with redshift. Since the whole population of disrupted dwarfs is composed of systems accreted at different redshifts, it is interesting to quantify the level of evolution of the MZR during different cosmic epochs, and to connect it to the scatter observed in the [Fe/H] -- $M_*$ distribution of disrupted dwarfs. For this analysis, we include the entire population of disrupted dwarfs in the simulations. Specifically, for each progenitor, we compute the $M_*$ and mean [Fe/H] values at accretion ($z_{\rm acc}$). After this, the disrupted dwarfs are separated into different  $z_{\rm acc}$ ranges ($[2.0,\,3.5]$, $[1.5,\,2.0]$, $[0.5,\,1.5]$ and $[0.0,\,0.5]$) and, for each range, we compute the mean [Fe/H] values in different $M_*$ bins. The mean MZRs obtained from this procedure are shown in Fig.~\ref{fig:Disrupted-MZR-different-z-single-panel}. (We note that the results are similar if medians are used instead). 
 
Fig.~\ref{fig:Disrupted-MZR-different-z-single-panel} shows that the normalization of the MZR evolves with redshift, changing by $\approx 0.25$~dex between the $z_{\rm acc}$ bins chosen here. As expected, the average [Fe/H] of progenitors increases with time, given that systems accreted later had more time to form stars and enrich in metals  prior to their accretion onto the host. Also, there does not seem to be a marked difference between the latest accreted dwarfs that are now disrupted (i.e., the lowest $z_{\rm acc}$ bin) and those that survive  (grey curve), the two populations overlapping. This is because both of these populations are mainly composed of high mass dwarfs, which are more metal-rich.

This figure also shows that the slopes of the MZRs of the disrupted populations  accreted at different $z_{\rm acc}$ are very similar, with values consistently around $0.27-0.29$. This suggests that the slope of the MZR of the MW progenitors does not change with redshift (only its normalization). Also, the slopes for the early progenitors are similar with the slope for surviving satellites (which is $\simeq 0.32$ in our simulations, see the grey line in  Fig.~\ref{fig:Disrupted-MZR-different-z-single-panel}; and $\simeq 0.3$ in the Local Group).

Overall, the slope of the combined population of disrupted dwarfs (of 0.48; see eq.~\ref{eq:MZRfit_d} and the translucent purple line in Fig.~\ref{fig:Disrupted-MZR-different-z-single-panel}) is steeper than the slopes for the individual $z_{\rm acc}$ bins. This can be explained by the increase in the MZR normalization with decreasing $z_{\rm acc}$, plus the changes in the stellar mass distributions of the disrupted dwarfs with $z_{\rm acc}$. The latter are shifting progressively towards higher masses at lower redshifts, and this is because later accreted dwarfs have more time to form stars prior to accretion. 

Finally, as discussed in \citet{Naidu2022}, the disrupted population in the MW includes systems accreted at intermediate redshifts ($z_{\rm acc} \approx 0.5-2$). In the case of \texttt{ARTEMIS}, Fig.\ref{fig:Disrupted-MZR-different-z-single-panel} shows that, if we were to limit the sample of disrupted galaxies to this redshift interval, their mean MZR slope would be very similar to that derived for the simulated surviving population, in very good agreement with \citet{Naidu2022} findings. A discussion about the evolution of MZR with redshift, in the context of other studies, is presented in Section~\ref{sec:discussion}.

\subsection{[Mg/Fe] -- stellar mass and [Mg/Fe] -- [Fe/H] distributions} 
\label{ssec:alpha}

Similar to the analysis of the [Fe/H] distributions of the disrupted and surviving dwarf populations, we investigate the different trends in terms of their alpha-enhancements.

In the left panel of Fig.~\ref{fig:alpha-vs-Mstar} we show the [Mg/Fe] -- ${M}_*$ distributions in the combined disrupted and surviving populations in the simulations (purple stars and grey squares, respectively), compared with the corresponding distributions in the MW populations (data are again from \citealt{Naidu2022}). The right panel of the same figure shows the corresponding [Mg/Fe] -- [Fe/H] distributions. 

Generally, the [Mg/Fe] abundances decrease with increasing stellar mass, for both surviving and disrupted satellites. This is because more massive galaxies sustain star formation for longer, and enrich more in Fe from Type Ia supernovae. Lower mass dwarfs have shorter periods of star formation, and therefore they enrich mainly in $\alpha$-elements (namely, Mg), which are created in Type II supernovae that occur on shorter timescales than Type Ia \citep{Tinsley1979}. We note that the time of accretion is also an implicit factor here, as later accreted dwarfs have more extended star formation histories, and therefore have higher [Fe/H] and lower [Mg/Fe] at accretion. The dependency of chemical abundances on $z_{\rm acc}$ will be investigated in more detail in the next section.

Overall, the [Mg/Fe] distributions show a plateau (or a shallow slope) at the metal-poor end ([Fe/H] $\lesssim -1$), followed by a decrease towards higher metallicities. This trend is seen in both disrupted and surviving populations, and  for both simulated dwarfs and observed ones. Also, the [Mg/Fe] -- [Fe/H] trend obtained for the disrupted dwarfs is similar to the trend observed in the stellar halo of the MW (e.g., \citealt{Venn2004}), confirming the ability of  $\Lambda$CDM models to produce realistic (accreted) stellar haloes of MW-mass galaxies.

One main result in Fig.~\ref{fig:alpha-vs-Mstar} is that simulations predict disrupted populations that are more enriched in [Mg/Fe] than the surviving ones, at both fixed $M_*$ (left panel), and at fixed [Fe/H] (right panel). The trend agrees with the one seen in [Mg/Fe] between corresponding populations in the MW. Note, however, that the simulated chemical abundances show a large scatter in these offsets, compared to the MW data. This is mainly due to the inclusion of populations from all 42 simulated MW-mass systems. If, instead, we analyse the offsets on a system-by-system basis (as shown in Appendix \ref{sec:s_6}), their magnitudes are similar to the one measured between the MW populations. 

In conclusion, the simulations are in qualitatively good agreement with the MW observations, in that: {\it i)} the observed [Fe/H] and [Mg/Fe] values for surviving and disrupted dwarfs in the MW are matched by the distribution of corresponding values in the simulations; {\it ii)} the simulations retrieve the general trends seen in the MW observations, for example, the decrease in [Mg/Fe] with $M_*$, or the typical [Mg/Fe] -- [Fe/H] ``chemical clock'' (e.g., \citealt{Matteucci2021}); {\it iii)} the simulations also predict that the disrupted dwarfs are more enhanced in [Mg/Fe] than the surviving ones, both at fixed ${M}_*$ and at fixed [Fe/H]; {\it iv)} in addition, the simulations match the observed MZRs for the two populations and retrieve the observed offset in [Fe/H] between disrupted dwarfs and surviving dwarfs at fixed $M_*$ (Fig.~\ref{fig:MZR-scatter-fits}).

\subsection{The scatter in the chemical abundance versus stellar mass}
\label{ssec:scatter}

The nature of the scatter in the simulated distributions deserves further investigation. Figs.~\ref{fig:MZR-scatter-fits} and ~\ref{fig:alpha-vs-Mstar} show significant scatter in the chemical abundances of both disrupted and surviving populations. These figures also highlight two points: 1) the scatter is larger in the disrupted population than in the surviving one; and 2) the simulated chemical abundances present a larger scatter in both populations, compared with observations of disrupted and surviving satellites in the MW.

The latter point is plausibly explained by the inclusion of populations from many more MW-mass systems in the simulations (42), compared with the single system in the observations. The larger scatter in the disrupted populations compared with the surviving ones can also be explained, in part, by the larger number of disrupted systems over the lifetime of a MW-mass galaxy \citep{Fattahi2020} compared with the number of surviving dwarfs, of which there are typically a dozen or so per host halo (see \citealt{Font2021}). 

However, some of the scatter in the disrupted populations may be due to different accretion histories of the simulated MW-mass systems (e.g., different combinations of stellar masses and/or different accretion times in each disrupted dwarf population). In particular, the large variation in chemical abundances (both [Fe/H] and [Mg/Fe], at fixed $M_*$) in the disrupted populations suggests a connection with the accretion histories. For example, at fixed stellar mass satellites that were accreted relatively earlier will typically have had higher star formation rates than satellites of the same stellar mass but which were accreted later on. This will lead to different chemical abundances at the time of accretion. In addition, the star formation histories of satellites can be affected by the environmental conditions specific to their hosts; for example, satellites that have spent more time in denser environments (i.e., closer to their hosts) may have had their star formation rates curtailed which in turn will affect their chemical evolution.

\begin{figure*}
\includegraphics[width=1.6\columnwidth]{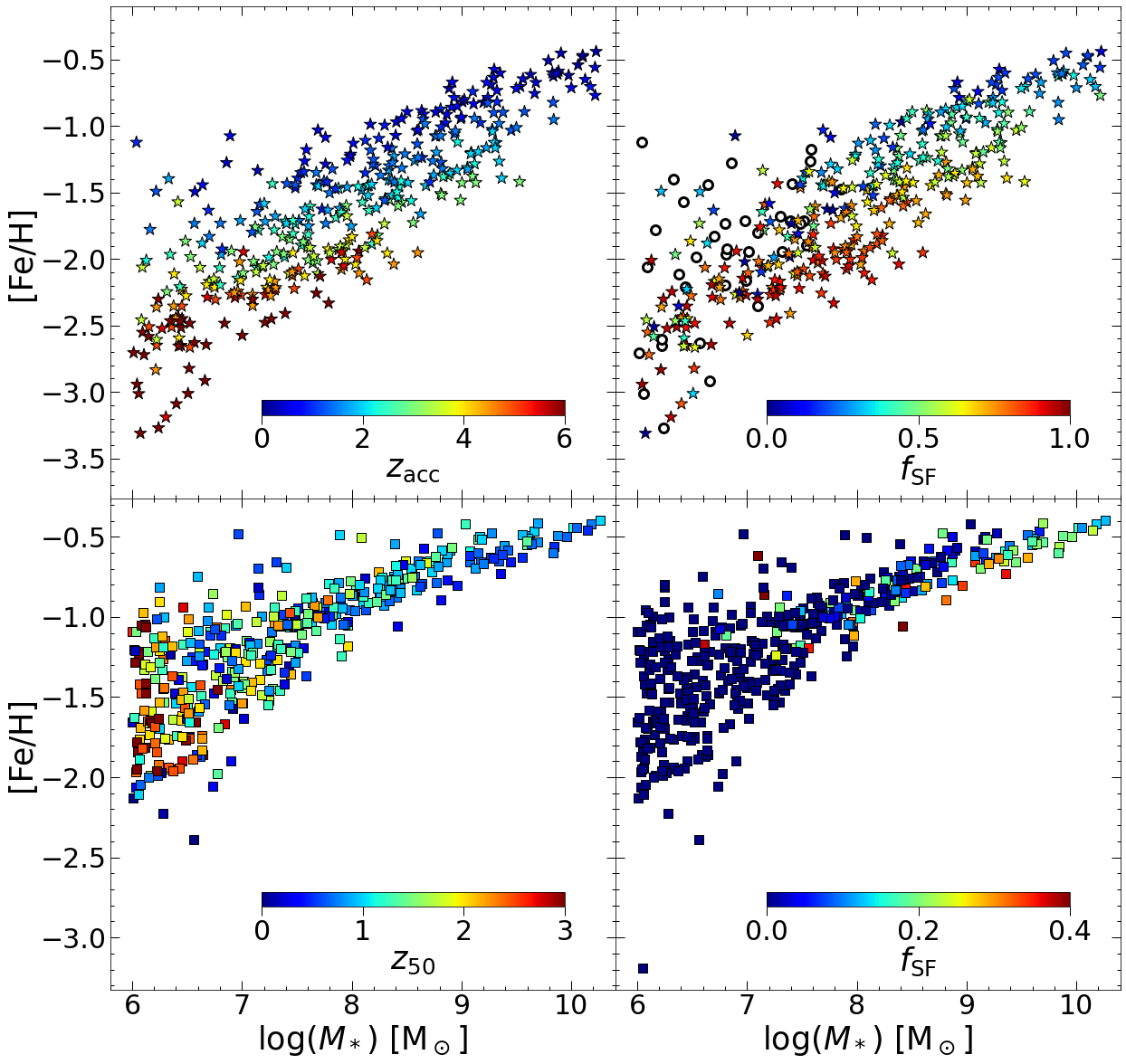}
\caption{The [Fe/H] -- $M_*$ distributions of the disrupted {(\it top panels)} and surviving {(\it bottom panels)} populations. The symbols for the disrupted dwarfs are colour-coded by the redshift of accretion, $z_{\rm acc}$ ({\it left}) and by the star-forming gas fraction, $f_{\rm SF}$ ({\it right}), while for the surviving ones, they are colour-coded by the redshift of reaching half of maximum stellar mass, $z_{\rm{50}}$ ({\it left}) and by $f_{\rm SF}$ ({\it right}), respectively. Mean [Fe/H], $M_*$ and $f_{\rm SF}$ values are computed at $z_{\rm acc}$ for the disrupted satellites, and at $z=0$ for the surviving ones, respectively. A small population of disrupted dwarfs has no star-forming gas at accretion, $f_{\rm{SF}}=0$ (empty circles). For the disrupted dwarfs, a clear correlation is seen between the scatter in [Fe/H] at fixed $M_*$ and $z_{\rm acc}$ or $f_{\rm SF}$. Correlations are very weak for the surviving populations.}
\label{fig:MZR-z50-fSF}
\end{figure*}

\begin{figure*}
\includegraphics[width=1.6\columnwidth]{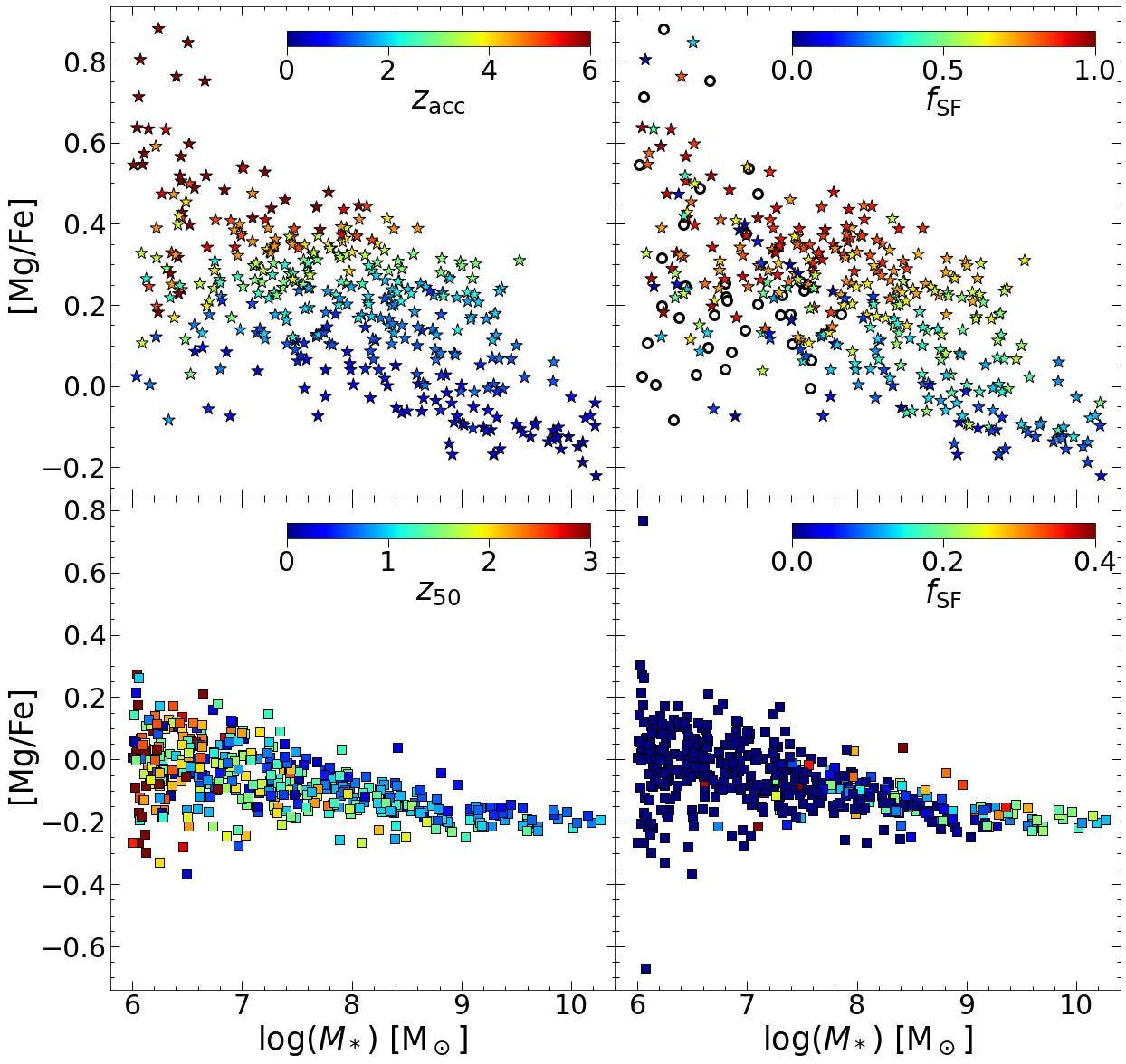}
\caption{{\it Top:} [Mg/Fe] -- ${M}_*$ distributions for the disrupted populations, with values colour-coded by $z_{\rm acc}$ ({\it left}) and by $f_{\rm SF}$ at accretion ({\it right}). Systems with $f_{\rm SF}=0$ are shown with empty circles. A clear correlation of the scatter in [Mg/Fe] at fixed $M_*$ is seen for the disrupted dwarfs, with both $z_{\rm acc}$ and $f_{\rm SF}$. {\it Bottom:} The same distributions for the surviving dwarfs, colour-coded by $z_{\rm{50}}$ ({\it left}) and by $f_{\rm SF}$ at $z=0$ ({\it right}). The correlations of the scatter with these parameters are much weaker in this case.}
\label{fig:alpha-Mstar-z50-fSF}
\end{figure*}

To test this scenario, and determine how different accretion events can be identified in the scatter of the [Fe/H] -- $M_*$ and [Mg/Fe] -- $M_*$ distributions, we choose a few simple parameters as proxies for the accretion histories. First, we use a redshift associated with accretion or the mass growth of each dwarf, specifically, $z_{\rm acc}$ for the disrupted dwarfs and $z_{50}$ for the surviving ones (as discussed in Section ~\ref{sec:merger_trees}). We also use the cold, star forming (SF) gas fraction in each dwarf satellite ($f_{\rm SF}$), as this parameter has previously been found to correlate well with the distribution of chemical abundances (for example, with the metallicity scaling relations; see \citealt{DeRossi2017, DeRossi2018, Maiolino2019} and references therein).  

Fig.~\ref{fig:MZR-z50-fSF} shows the [Fe/H] -- ${M}_{*}$ distribution for the  disrupted and surviving dwarf populations in the simulations, using the same samples as in the left panel of Fig.~\ref{fig:MZR-scatter-fits}. As before, [Fe/H] are the means of the stellar [Fe/H] values in these dwarfs, measured at $z_{\rm acc}$ for disrupted dwarfs, and at $z=0$, for the surviving ones. For the disrupted populations (top panels), the distributions are colour-coded by $z_{\rm acc}$ (left) and $f_{\rm SF}$ (right), while for the surviving ones (bottom panels), are colour-coded by $z_{50}$ (left) and $f_{\rm SF}$ (right), respectively.  The cold SF gas fraction is defined as usual (see \citealt{DeRossi2017}): $f_{\rm SF} = {M}_{\rm SF} / ({M}_* + {M}_{\rm SF})$, where ${M}_{\rm SF}$ is the total gas mass in a dwarf galaxy which fulfills the required conditions to form new stars (see \citealt{Schaye2015}). Note that $f_{\rm SF}$ are computed at $z_{\rm acc}$ for the disrupted dwarfs, but at $z=0$ for the surviving ones. (As it was done for Fig.~\ref{fig:MZR-scatter-fits}, the values for the surviving populations are computed at $z=0$ to facilitate the comparison with observations).

This figure shows that, for the disrupted dwarfs at least, the scatter is strongly dependent on both $z_{\rm acc}$ and $f_{\rm SF}$. At fixed ${M}_*$, dwarfs that are accreted early are more metal-poor; also, at fixed $M_*$, dwarfs that are more metal-poor are also more gas rich.  The correlation between $z_{\rm acc}$ and $f_{\rm SF}$ is not surprising, given that these parameters are not independent of each other. Dwarfs accreted early are expected to be more gas-rich, first because gas fractions generally were higher given that there has been less cosmic time to form stars; also, early on, dwarf galaxies can still retain  their gas reservoir, whereas later, this gas gets depleted through supernovae winds and ongoing star formation. 

The magnitude of the scatter in [Fe/H] also appears to vary with $M_*$, the lower mass dwarfs showing the largest amount of scatter, and highest mass dwarfs, the least. In general, the magnitude of the scatter correlates with the range in $z_{\rm acc}$ and $f_{\rm SF}$ values. For example, disrupted progenitors of $M_* \approx 10^6 - 10^7\, {\rm M}_{\odot}$ are accreted at redshifts ranging from $z_{\rm acc} \approx 6$ to 0, whereas those of $M_* \approx 10^9 - 10^{10}\, {\rm M}_{\odot}$ are accreted from $z_{\rm acc} \approx 1$ to 0. This is expected, as low mass dwarfs tend to form first and they could be accreted at any redshift, whereas high mass dwarfs require more time to grow and therefore are typically accreted late. A similar behaviour is seen in terms of $f_{\rm SF}$: low mass progenitors displaying a wide range of $f_{\rm SF}$ values, corresponding to systems that are either gas-rich and star-forming, or that are partially quenched, whereas all LMC-mass systems have low $f_{\rm SF}$ values, which indicates that they consumed most of their cold gas by the time of accretion. There is also an extreme case of progenitors without any cold gas at accretion ($f_{\rm SF}=0)$, which are indicated with empty circles in the top right panel of Fig~\ref{fig:MZR-z50-fSF}. These progenitors tend to be of very low mass, so they are prone to having their gas stripped by environmental processes, or to have their gas expelled by supernovae winds. We note, however, that towards the very low mass end, some of the scatter in dwarf properties may be due to the limited numerical resolution of the simulations.  

The change in the magnitude of the scatter with $M_*$ is also seen in the surviving population (bottom panels of Fig.~\ref{fig:MZR-z50-fSF}), however in this case the correlations with $z_{50}$ or $f_{\rm SF}$ are much weaker. Low mass surviving dwarfs still form at a wide range of redshifts  ($z_{50}\approx 3 -0$), and high mass ones typically form late ($z_{50}\approx 1-0$). Overall, however, the range in $z_{50}$ (and, by extension, in $z_{\rm acc}$) in the surviving dwarfs is narrower than the $z_{\rm acc}$ range in the disrupted dwarfs, both at fixed $M_*$ and across all masses. This is because the surviving dwarfs are typically accreted late. Moreover, there does not seem to be a clear correlation between $z_{50}$ and [Fe/H] of the surviving dwarfs at fixed $M_*$, as it was previously the case between $z_{\rm acc}$ and [Fe/H], for the disrupted ones.  The dependence on $z_{50}$ is more pronounced across the range of stellar masses, with progenitors that form early being typically more metal-poor.

Similarly, the present-day $f_{\rm SF}$ fractions in the surviving populations do not correlate with the scatter in [Fe/H] at fixed $M_*$. There is also very little variation in $f_{\rm SF}$ values across the range in dwarf stellar masses, most surviving dwarfs being almost devoid of cold gas at present time. This indicates that this population has been efficiently quenched by the environment of their host (this result is also in agreement with the findings of \citet{Font2022}, and is in agreement with observations of quenched fractions of satellites in MW and other MW analogues{\color{red} \citep{Geha2012,Mao2021,Wetzel2012,Greene2023,Karunakaran2023}} It is only the most massive ones (with $M_* \simeq 10^9 - 10^{10}\, {\rm M}_{\odot}$) that are able to retain some of their cold gas and still form stars today. These results can be explained by surviving populations being accreted typically late ($z_{50} \lesssim 1$), and therefore they tend to be similar in their physical properties. By consequence, this population bears little resemblance to the past accretion histories of their host.

Fig.~\ref{fig:alpha-Mstar-z50-fSF} shows a similar analysis of the scatter for the [Mg/Fe] -- ${M}_*$ distributions in the two dwarf populations. For the disrupted population (top panels), there is a considerable scatter in [Mg/Fe], particularly for systems with $M_* \lesssim 10^9\, {\rm M}_{\odot}$ (the region where [Mg/Fe] values exhibit a plateau, or a very slow decrease with increasing $M_*$). Above this mass, the [Mg/Fe] values of progenitors decrease more steeply with with $M_*$, and the scatter is also reduced. This is the regime of later accreted progenitors, which, due to their continued star formation, evolved chemically to lower [Mg/Fe] values. 

For the early accreted progenitors of disrupted dwarfs, the scatter in [Mg/Fe] values correlates, at fixed ${M}_*$, with both $z_{\rm acc}$ and with $f_{\rm SF}$. For progenitors of the same stellar mass, those accreted earlier have higher [Mg/Fe] abundances than those accreted late. Likewise, those progenitors with higher [Mg/Fe] at accretion, tend also to have higher gas fractions. However, the correlation of the scatter with $z_{\rm acc}$ at fixed $M_*$ (left panel), appears to be stronger than the correlation with $f_{\rm SF}$ values (right panel). This suggests that, although $z_{\rm acc}$ and $f_{\rm SF}$ are themselves correlated, they do not always track each other. The $f_{\rm SF}$ values appear to be either very high, or very low, which indicates that the progenitors may experience quenching processes on a relatively short timescales.

We note that some of the low-mass, metal-poor disrupted satellites present unusually high [Mg/Fe] abundances, e.g., $>0.4$. These galaxies were accreted at high redshifts ($z>4$, see Section~\ref{ssec:scatter}) which may indicate that the nucleosynthetic yield tables implemented in the \texttt{ARTEMIS} simulations are more uncertain at these ranges. The choice of the time delay between SNII and SNIa may also contribute to unusually high [Mg/Fe] abundances for the early accreted dwarfs. $\alpha$ elements are mainly synthesized in SNII, whereas Fe are mainly produced in SNIa. Since SNII occur in younger stellar populations than SNIa, a high level of $\alpha$-enhancement in early accreted satellites is expected. \citet{Wiersma2009} found that, up to $10^{8}~\rm{yr}$, the integrated metal ejecta in their model is dominated by SNII, whereas the contribution of SNIa for Fe becomes important after $10^9~\rm{yr}$.

The bottom panels in Fig.~\ref{fig:alpha-Mstar-z50-fSF} show the [Mg/Fe] -- ${M}_{*}$ distributions for the surviving dwarfs, colour-coded by $z_{50}$ and by $f_{\rm{SF}}$, respectively. A weak correlation with $z_{50}$ is seen at the low-mass end, where at  fixed $M_*$, dwarfs with higher [Mg/Fe] tend grow in mass more recently than those with lower [Mg/Fe]. This behaviour is not seen at higher masses though. At low masses, surviving dwarfs with higher $z_{50}$ have more time to evolve, increasing their Fe content and therefore having lower [Mg/Fe]. The right panel shows that, for massive systems (${M}_* \ga 10^{7}~\rm{M_{\sun}}$) and at fixed ${\rm M}_*$, dwarfs with higher [Mg/Fe] tend to have higher fractions of star-forming gas.

\begin{figure}
\includegraphics[width=0.95\columnwidth]{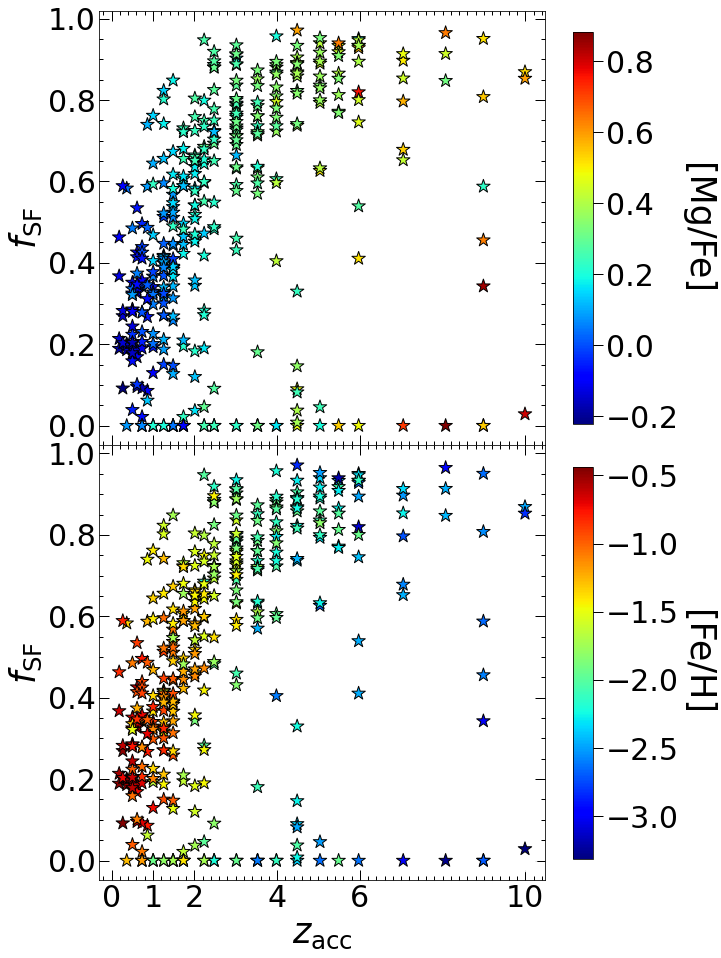}
\caption{Star-forming gas fraction, $f_{\rm SF}$, versus redshift of accretion, $z_{\rm acc}$, for disrupted dwarfs. Values are colour-coded by [Mg/Fe] ({\it top}) and by [Fe/H] ({\it bottom}), respectively. The $f_{\rm{SF}}$ values tend to decrease with redshift, with a sharp turnover around $z_{\rm{acc}} \approx 4$. Before this redshift, there is a considerable scatter in the relation between these parameters, however after this redshift the two parameters correlate well.}
\label{fig:disrupted-fSFvsalpha-cmap}
\end{figure}

We have seen that, for the disrupted dwarfs, $z_{\rm{acc}}$ and the fraction of cold gas these systems have at this redshift, $f_{\rm{SF}}$ are correlated. Typically, progenitors accreted earlier are more gas rich. We explore this correlation in more detail in Fig.~\ref{fig:disrupted-fSFvsalpha-cmap}, where we plot $f_{\rm{SF}}$ versus $z_{\rm{acc}}$ for all disrupted dwarfs, with star symbols colour-coded by the mean chemical abundances of each dwarf ([Mg/Fe] and [Fe/H], respectively), measured at accretion. Generally, $f_{\rm SF}$ values are high at early times and decrease sharply below $z_{\rm acc} \simeq 4$. The scatter in the relation between $f_{\rm SF}$ and $z_{\rm acc}$ is fairly large above redshift $\approx 4$, however towards low redshift the two parameters correlate well. The galaxies  lower $f_{\rm{SF}}$ values than the average could be related to environmental effects or stochastic star formation, particularly at the low mass end. Additionally, the $f_{\rm SF}$ values could be affected by numerical resolution. For example, the population of low-mass dwarfs with $f_{\rm SF} \simeq 0$  (shown with empty circles in Figs.~\ref{fig:MZR-z50-fSF} and \ref{fig:alpha-Mstar-z50-fSF}) display a wide range of $z_{\rm acc}$ values.  This suggests that their gas has been removed/consumed at a much faster pace than the average population accreted at that redshift. Note that the apparent vertical stripes in $z_{\rm acc}$, which are most visible at high redshifts, are just due to frequency of simulation snapshots that have been output.  The merger tree algorithm does not interpolate between snapshots but simply uses the redshift of the snapshot which is nearest to the accretion event.

In summary, the scatter in averaged chemical abundances ([Fe/H] and [Mg/Fe]) of disrupted dwarfs at fixed $M_*$ is larger than the corresponding scatter in the surviving dwarfs. Also, the scatter in the distributions for the disrupted dwarfs correlates with the $z_{\rm acc}$ and with the cold gas fraction at the time of accretion, $f_{\rm SF}$ of the progenitors, whereas in the case of surviving dwarfs, there is no correlation in the scatter with $z_{50}$ (or $z_{\rm acc}$), and their present-day $f_{\rm SF}$ fractions are typically very low. Also, the chemical abundances of the disrupted dwarfs are more diverse than those of the surviving satellites, which are similar regardless of the properties of their host (e.g., total mass) or the past accretion history in that system. This suggests that the chemical abundances of disrupted population may be a good indicator of the accretion history of their host, for example using chemical abundances to extract the stellar masses and redshifts of accretion of the progenitors. This may help in the reconstruction of the formation histories of MW-mass galaxies, as chemical abundances are more readily available from the debris found in the MW today, rather than total stellar masses and accretion redshifts, which usually are inferred via  modelling of the orbits of satellite progenitors.

\subsection{Global properties of the disrupted dwarfs populations} 
\label{ssec:av_prop}

\begin{figure}
\includegraphics[width=0.95\columnwidth]{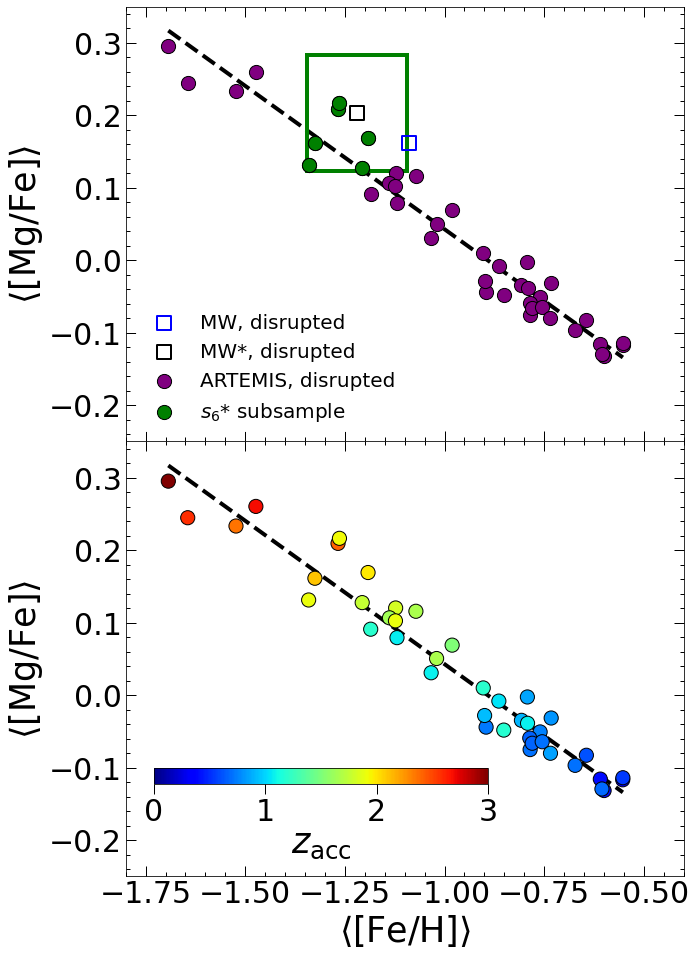}
\caption{{\textit{Top:}} The $\langle {\rm [Mg/Fe]} \rangle$ -- $\langle {\rm [Fe/H]} \rangle$ relation of disrupted populations in the simulations (filled circles). Symbols represent the mass averages of each disrupted population in the 42 simulated MW-mass systems. $\langle {\rm [Mg/Fe]} \rangle$ anti-correlates with $\langle {\rm [Fe/H]} \rangle$. The best fit is a linear relation with a slope of $-0.40\pm0.015$. The blue and black squares correspond to the samples of disrupted dwarfs in the MW: including I'itoi and Sagittarius (MW sample) and excluding them (MW* sample), respectively. The six closest populations to the MW* data in this relation are shown with green circles and called the ``$s_6$* subsample.'' {\textit{Bottom:}} Same relation as above, colour-coded by the $\langle z_{\rm{acc}} \rangle$ of each disrupted population. There is a strong anti-correlation between $\langle z_{\rm{acc}} \rangle$ and $\langle {\rm [Fe/H]} \rangle$ (recently accreted populations are more metal-rich) and a strong correlation between $\langle z_{\rm{acc}} \rangle$ and $\langle {\rm [Mg/Fe]} \rangle$ (recently accreted populations have lower [Mg/Fe] abundances).} 
\label{fig:Weighted_alpha_vs_Z_zacc_cmap}
\end{figure}

Our results so far show that the chemical abundances of the disrupted dwarfs correlate with their $z_{\rm acc}$, suggesting that the accretion history of a MW-mass galaxy could be inferred from the chemical abundance properties of its disrupted population. In contrast, the chemical abundance properties of surviving populations (see bottom panels of Figs.~\ref{fig:MZR-z50-fSF} and \ref{fig:alpha-Mstar-z50-fSF}), seem to have little connection with either the properties of their hosts, or with those of the disrupted population. 

In this section we explore the global properties of the disrupted populations, in order to determine which diagnostics could be used to infer the accretion histories of their hosts.
Specifically, we select several observables that involve the chemical abundance distributions of the disrupted populations. These are: the average slopes of the MZRs of the disrupted populations, their mean [Fe/H] and [Mg/Fe] abundances at accretion, and the average slopes of the [Mg/Fe] -- ${M}_*$ distributions. These diagnostics are motivated by the correlation with $z_{\rm acc}$ in the scatter of these relations for the disrupted populations (see Figs.~\ref{fig:Disrupted-MZR-different-z-single-panel},~\ref{fig:MZR-z50-fSF} and \ref{fig:alpha-Mstar-z50-fSF}).  The accretion histories are also measured globally, using the average redshift of accretion of the disrupted population, $\langle z_{\rm acc} \rangle$, which is defined as the stellar mass-weighted mean of individual $z_{\rm acc}$ of all disrupted dwarfs in a given MW-mass host. Likewise, for chemical abundances, we compute mass-weighted, global values for the entire population of disrupted dwarfs in a given MW-mass system. Therefore,

 \begin{equation}
 \langle X \rangle=\sum_{i} \frac{{M}_{*,i} X_i}{{M}_{*,i}},
\label{eq:weighted_values}
 \end{equation}

\noindent where the index $i$ denotes a given disrupted dwarf galaxy, ${M}_{*,i}$ is its stellar mass measure at accretion, and $X$ is the parameter that is being computed (i.e., $z_{\rm acc}$, [Fe/H] or [Mg/Fe]). 

We compute the same global parameters  $\langle X \rangle$ for the disrupted dwarfs in the MW,  using the data from \citet{Naidu2022}. We note that these authors use a different definition for $z_{\rm acc}$, namely as the redshift  associated with the first pericentric passage of the progenitor of the disrupted dwarf inside the MW. We retain this definition in our computation of $\langle z_{\rm acc} \rangle$ for the observed MW population, however we note that in some cases the progenitors of the disrupted dwarfs can spend a significant amount time between accretion onto the host and their first pericentric passage inside it (or the final time of disruption), as will be discussed in Section~\ref{ssec:disruption}.

Fig.~\ref{fig:Weighted_alpha_vs_Z_zacc_cmap} shows the relation between the mean chemical abundance values, $\langle {\rm [Mg/Fe]} \rangle$ and $\langle {\rm [Fe/H]} \rangle$, for the disrupted populations in all simulated MW-mass systems. Each full circle corresponds to a disrupted population of a given MW-mass system. To these, we add the corresponding values for the disrupted population in the MW. As before, we use data from \citet{Naidu2022}, and consider two cases: one in which Sagittarius and I'itoi are included in the sample (blue square), and one in which they are not (black square). Hereafter, the first sample (which contains 9 dwarfs) is called ``MW", whereas the second sample (containing 7 dwarfs) is called ``MW*''. As discussed in Section~\ref{ssec:MZR}, I'itoi and Sagittarius were excluded by \citet{Naidu2022} in their estimate of the MZR for the disrupted population, however we find that by including these two systems we obtain a better agreement between the predicted MZRs of disrupted populations and the observations (i.e., the ``MW full sample'' in the left panel of Fig.~\ref{fig:MZR-scatter-fits}). 

This figure shows a strong anti-correlation between the global $\langle {\rm [Mg/Fe]} \rangle$ and $\langle {\rm [Fe/H]} \rangle$ values of the simulated disrupted populations. The best linear fit (shown with dashed line in both panels) has a slope of $-0.40\pm0.015$, with a corresponding Spearman coefficient of $r_s=-0.95$. Moreover, both observational data points (both the MW and MW* samples) follow the predicted $\langle {\rm [Mg/Fe]} \rangle$ -- $\langle {\rm [Fe/H]} \rangle$ relation. The bottom panel of Fig.~\ref{fig:Weighted_alpha_vs_Z_zacc_cmap} shows the same relation for the simulated disrupted populations as the top panel, but now colour-coded by  $\langle z_{\rm{acc}} \rangle$ (computed as in eq.~\ref{eq:weighted_values}. This shows an anti-correlation between $\langle {\rm [Fe/H]} \rangle$ and $\langle z_{\rm{acc}} \rangle$ and a correlation between $\langle z_{\rm{acc}} \rangle$ and $\langle {\rm [Mg/Fe]} \rangle$. These resemble the individual relations between [Fe/H] and $z_{\rm acc}$, and between [Mg/Fe] and $z_{\rm acc}$ for disrupted dwarfs seen in Figs.~\ref{fig:MZR-z50-fSF} and \ref{fig:alpha-Mstar-z50-fSF}, but in this case, the correlations are much stronger.

Overall, these relations between $\langle {\rm [Mg/Fe]} \rangle$ and $\langle {\rm [Fe/H]} \rangle$, and each in turn with $\langle z_{\rm{acc}} \rangle$, suggest that these observables are good diagnostics for constraining the accretion histories of MW-mass hosts. However, we note that as these are global parameters of a given disrupted population, they are more sensitive to the most massive accretion events in this population (given that they are mass-weighted estimates of the accretion history).

For a more in-depth comparison with the disrupted population of the MW, we choose six simulated disrupted populations whose $\langle {\rm [Mg/Fe]} \rangle$ and $\langle {\rm [Fe/H]} \rangle$ values fall close to the measurements in the MW* sample. These populations are hereafter called  the ``$s_6$* subsample''.  These are located inside the green rectangle centred on the MW* data point in the top panel of Fig.~\ref{fig:Weighted_alpha_vs_Z_zacc_cmap}, and are also shown with green circles. In the \texttt{ARTEMIS} simulations,  these populations correspond to the MW-mass hosts: G1, G23, G26, G29, G38 and G42 (see \citealt{Font2020}). We note that four of these MW-mass hosts (G1, G23, G29 and G42) are among the disc galaxies with a GES feature identified by \citet{Dillamore2022}. For a detailed characterization on a system-by-system basis in the $s_6$* subsample, see Appendix~\ref{sec:s_6}. We prefer to select our subsample based in its relation to the MW* sample instead of the MW one since the analysis carried out by \citet{Naidu2022} is mainly done on the former.

\begin{figure}
\includegraphics[width=0.95\columnwidth]{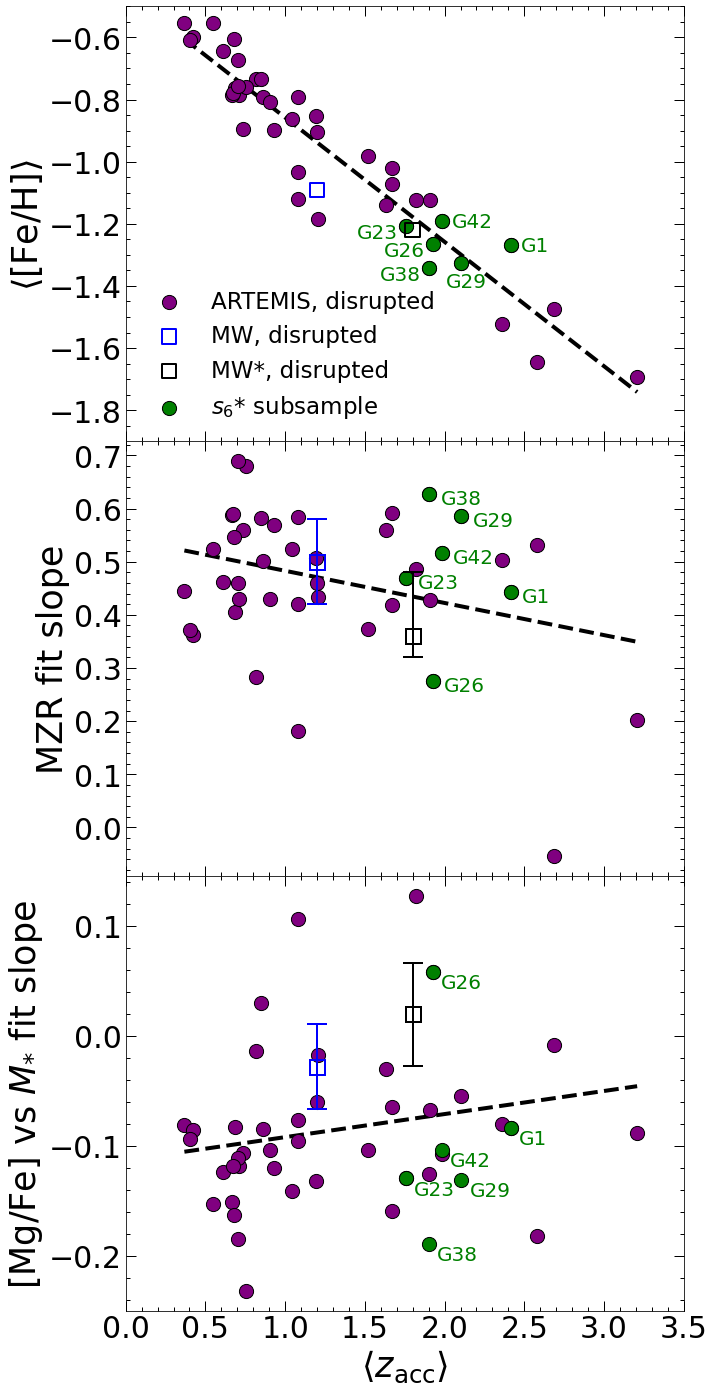}
\caption{Scaling relations between $\langle z_{\rm acc} \rangle$ and global parameters of disrupted dwarf populations in the simulated MW-mass hosts. {\it Top:} $\langle {\rm [Fe/H]} \rangle$--$\langle z_{\rm acc} \rangle$ relation.  {\it Middle:}  the MZR slope versus $\langle z_{\rm{acc}} \rangle$. {\it Bottom:} the slope of the [Mg/Fe] -- ${M}_{*}$ relation versus $\langle z_{\rm{acc}} \rangle$.
Green circles correspond to the disrupted populations in the $s_6$* subsample, while purple circles correspond to those of the rest of the simulated populations. Values inferred from the MW disrupted population, both including and excluding I'itoi and Sagittarius, are shown with blue and black squares, respectively.
The black dashed lines represent the linear fits for the simulated relations (from top to bottom, the corresponding $r_s$ values are: $-0.95$, $-0.11$ and $0.30$, respectively).}  
\label{fig:Sl.opes_and_Z_vs_weighted_zacc}
\end{figure}

As the $s_6$* populations match the averaged properties of the MW* sample (in $\langle {\rm [Fe/H]} \rangle$, $\langle {\rm [Mg/Fe]} \rangle$ and $\langle z_{\rm acc} \rangle$), we want to know whether they also match other observables, for example the slope of the MZR in the MW* sample, or the slope of the [Mg/Fe] -- $M_*$ distribution. The latter are also potential useful diagnostics, as suggested by the results in previous sections.
The panels in Fig.~\ref{fig:Sl.opes_and_Z_vs_weighted_zacc} show, from top to bottom, with purple circles: the relation between $\langle {\rm [Fe/H]} \rangle$ and $\langle z_{\rm{acc}} \rangle$ for all disrupted populations in the simulations, the relation between the slopes of the MZR of these disrupted populations and the corresponding $\langle z_{\rm{acc}} \rangle$, and the relation between the slopes of the [Mg/Fe] -- $M_*$ relations of these populations and 
$\langle z_{\rm{acc}} \rangle$, respectively. The six disrupted populations in the $s_6$* subsample are highlighted again in green. The two disrupted population in the MW are shown again with empty squares, blue for the "MW" sample and black for "MW*". 

The top panel shows that $\langle {\rm [Fe/H]} \rangle$  correlates very strongly with $\langle z_{\rm{acc}} \rangle$. Our best linear regression fit to the $\langle {\rm [Fe/H]} \rangle$ -- $\langle z_{\rm{acc}} \rangle$ relation, shown in this panel with the black dashed line, is:

\begin{equation}
    \langle [\rm{Fe/H}] \rangle= (-0.40\pm 0.02) \, \langle z_{\rm{acc}} \rangle-(0.46\pm 0.03),
\label{eq:Fezacc}
\end{equation}

\noindent while the Spearman coefficient of this relation is $r_s=-0.95$. Interestingly, the same $r_s$ value was obtained for the $\langle {\rm [Mg/Fe]} \rangle$ -- $\langle {\rm [Fe/H]} \rangle$ relation in Fig.\ref{fig:Weighted_alpha_vs_Z_zacc_cmap}, which indicates that these diagnostics are equally strong for inferring the accretion history of the population. For example, one can infer the $\langle z_{\rm acc} \rangle$ of a given disrupted population by estimating both $\langle {\rm [Fe/H]} \rangle$  and $\langle {\rm [Mg/Fe]} \rangle$ and using the prediction from the bottom panel of Fig.~\ref{fig:Weighted_alpha_vs_Z_zacc_cmap}, or by estimating only $\langle {\rm [Fe/H]} \rangle$ and using the fit from eq.~\ref{eq:Fezacc}. These predictions are useful as they require only chemical abundance measurements, whereas estimating $\langle z_{\rm acc} \rangle$ from individual $z_{\rm acc}$ values typically would involve additional assumptions, e.g., about the total mass of the host, or about the orbits of the infalling dwarf galaxies.

The overall trend of increasing $\langle {\rm [Fe/H]} \rangle$ with decreasing $\langle z_{\rm acc} \rangle$ can be understood as a result of more chemical enrichment taking place in dwarf galaxies accreted late, as well as to the iron abundance being specially sensitive to the occurrence of delayed SNIa at later times. Overall, those disrupted dwarf populations which are accreted more recently tend to include more massive dwarfs, which are more metal-rich at the time of accretion. Note that this relation includes mass-weighted values, and so it tracks mainly the evolution of most massive disrupted dwarfs. The correlation found here is also expected on the basis of the most massive progenitors obeying a MZR. 

Our observational estimates for $\langle {\rm [Fe/H]} \rangle$  and $\langle z_{\rm{acc}} \rangle$ for the MW* and MW samples are consistent with the prediction from the simulations (eq.~\ref{eq:Fezacc}). In addition, the data points for the six disrupted populations in the $s_6$* subsample (green circles) fall in the vicinity of the MW* sample in the $\langle {\rm [Fe/H]} \rangle$  -- $\langle z_{\rm{acc}} \rangle$ plane (note that the $s_6$* subsample was selected on the basis of [Mg/Fe] and [Fe/H] alone), although their corresponding $\langle z_{\rm acc} \rangle$ values show a somewhat wide range, from 1.76 (G23) to 2.42 (G1). For the observed disrupted population (i.e., MW*), the simulated fit (eq.~\ref{eq:Fezacc}) predicts an average redshift of accretion of $\langle z_{\rm{acc}} \rangle \simeq 1.8$, which agrees with our estimate based on the MW's data from \citet{Naidu2022}. For the observational sample which includes both Sagittarius and I'itoi (MW sample, blue square), the simulated fit predicts a slightly lower value, of $\langle z_{\rm{acc}} \rangle$ $\simeq 1.6$, but the estimated value from the observations of $\simeq 1.2$ is within the simulated scatter at that $\langle {\rm [Fe/H]} \rangle$. The lower $\langle z_{\rm{acc}} \rangle$ / higher $\langle {\rm [Fe/H]} \rangle$ data point for the MW sample compared to MW* is due mainly to the inclusion of Sagittarius, which is accreted more recently and is also fairly massive. 

The middle panel of Fig.~\ref{fig:Sl.opes_and_Z_vs_weighted_zacc} shows distribution of the slopes of the MZRs in the simulated disrupted populations and the corresponding $\langle z_{\rm{acc}} \rangle$. (The individual MZRs of these populations have been shown before, in the right panel of Fig.~\ref{fig:MZR-scatter-fits}). The populations from the $s_6$* subsample are highlighted again in green, and the estimates for the observational MW* and MW samples are added with the same symbols as in the top panel.  The results do not support a correlation between the slope of the MZR and $\langle z_{\rm{acc}} \rangle$, which is consistent with the results in Fig.~\ref{fig:Disrupted-MZR-different-z-single-panel}. The linear regression fit to the simulated populations has Spearman coefficient of $r_s=-0.11$. The fit, shown with dashed black line in the middle panel, has a negligible slope of $-0.06\pm 0.03$ (the y-intercept is $0.54\pm 0.04$). This result is expected, as the individual MZRs of the disrupted populations are not significantly different in slopes and normalisations, although there is some scatter in these relations. The MZR slope estimates for the disrupted population in the MW are matched, within the scatter, by the simulations. 

The relation between the slope of the [Mg/Fe] -- ${M}_{*}$ relation of each disrupted population and the corresponding $\langle z_{\rm{acc}} \rangle$ is explored in the bottom panel of Fig.~\ref{fig:Sl.opes_and_Z_vs_weighted_zacc}. Here again, there is little or no correlation between the two parameters.
The linear regression fit to the simulated data (shown again with black dashed line) has 
a slope of $0.02\pm 0.02$ and a y-intercept of $-0.11\pm 0.02$, but a low Spearman coefficient, of $r_s=0.3$. 
Compared with this fit, the two slopes derived from observational data are high, but they are still within the predicted scatter. Out of the six disrupted populations in the $s_6$* subsample, G26 agrees most with the estimate for the MW* sample. 

In summary, simulations predict that the $\langle z_{\rm{acc}} \rangle$ of a disrupted population can be very well constrained by the mass-weighted chemical abundances ($\langle {\rm [Fe/H]} \rangle$ and $\langle {\rm [Mg/Fe]} \rangle$), but poorly constrained by other observables, such as the slope of the MZR of the population, or the slope of the [Mg/Fe] -- ${M}_{*}$ distribution. 

\begin{figure}
\includegraphics[width=0.9\columnwidth]{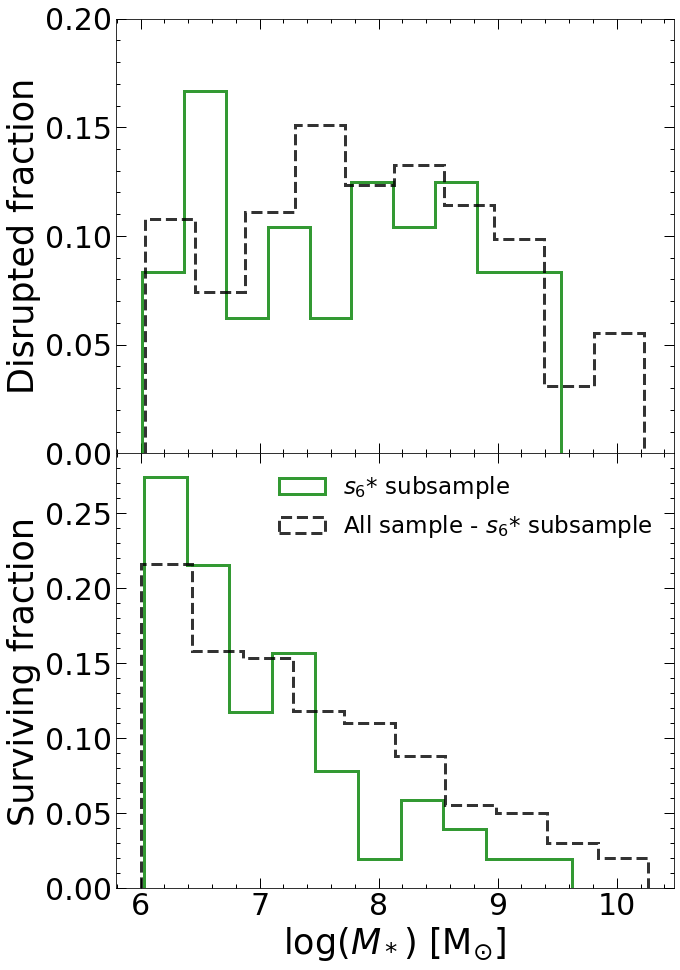}
\caption{{\it Top:} Stellar mass distribution of the progenitors of disrupted dwarfs in the $s_6$* subsample (solid green line) and in the remaining simulated hosts (black dashed line).  {\it Bottom:} Same distributions but for the surviving dwarf galaxies. The $s_6$* hosts have slightly higher fractions of lower mass progenitors than the rest of the sample, for both disrupted and surviving populations.}
\label{fig:Disrupted_mass_distribution}
\end{figure}

\subsection{The distribution of stellar masses in the disrupted and surviving populations}
\label{ssec:mass_dist}

Our analysis indicates that the disrupted progenitors tend to be increasingly more massive at accretion towards lower redshift (see, e.g., Fig.~\ref{fig:Disrupted-MZR-different-z-single-panel}). Here we investigate in more detail the distributions of stellar masses of the progenitors of disrupted and surviving dwarfs. We also track the progenitors of the six disrupted populations in the $s_6$* subsample separately, to assess whether the progenitors in the hosts resembling the MW (in terms of their global accretion histories) have a different stellar mass distribution from the populations in the rest of the simulated hosts.

In Fig.~\ref{fig:Disrupted_mass_distribution} the distributions of stellar masses of populations in the $s_6$* subsample are shown with solid green lines, while the corresponding distributions for the population in the rest of the simulated hosts are shown with black dashed lines. This is done separately for the disrupted and surviving populations (top and bottom panels, respectively). The disrupted populations (which, again, comprise systems accreted at various redshifts) show a more even distribution in stellar masses than the surviving populations. The stellar mass distribution of surviving satellites is similar to a power-law, which is the expected mass distribution of satellites at any given time. The roughly even distribution of masses in the disrupted population is the result of a combination of factors. Extending the period of accretion to effectively a Hubble time, means that stellar mass distribution encompasses more massive satellites. The more massive satellites are also more efficiently disrupted by tidal forces;  as the tidal forces scales as the square of the total mass of the satellite, this would lead to more massive satellites sinking faster and more likely to be fully disrupted (hence included in this plot). 

Comparing the two types of progenitors in the $s_6$* subsample versus those in the rest of the simulated hosts, we find that the corresponding stellar mass distributions are generally similar. There is however a slight tendency for the $s_6$* hosts to have more low-mass dwarf progenitors, for both disrupted and surviving populations. The trend is more noticeable for the surviving dwarfs, however this can be attributed to low number statistics (i.e., fewer number of hosts results in even more rare high mass, recent accretions).

The finding that disrupted populations in the $s_6$* subsample are composed of slightly less massive dwarfs than those in the rest of the hosts, is consistent with the slightly lower than average values of $\langle {\rm [Fe/H]} \rangle$ (and slightly higher than average $\langle {\rm [Mg/Fe]} \rangle$) values obtained for these populations with respect to the majority of the other MW-mass systems in Fig.~\ref{fig:Weighted_alpha_vs_Z_zacc_cmap}.  This is corroborated by the higher $\langle z_{\rm{acc}} \rangle$ of the disrupted populations in the $s_6$* subsample compared with the rest of the hosts in Fig.~\ref{fig:Sl.opes_and_Z_vs_weighted_zacc}, which is also consistent with a more significant population of lower mass dwarfs (as higher mass dwarfs require more time to form). This suggests that the $\langle z_{\rm{acc}} \rangle$ of a disrupted population is also indicative of its stellar mass spectrum, and vice versa.

\begin{figure}
\centering
\includegraphics[width=0.9\columnwidth]{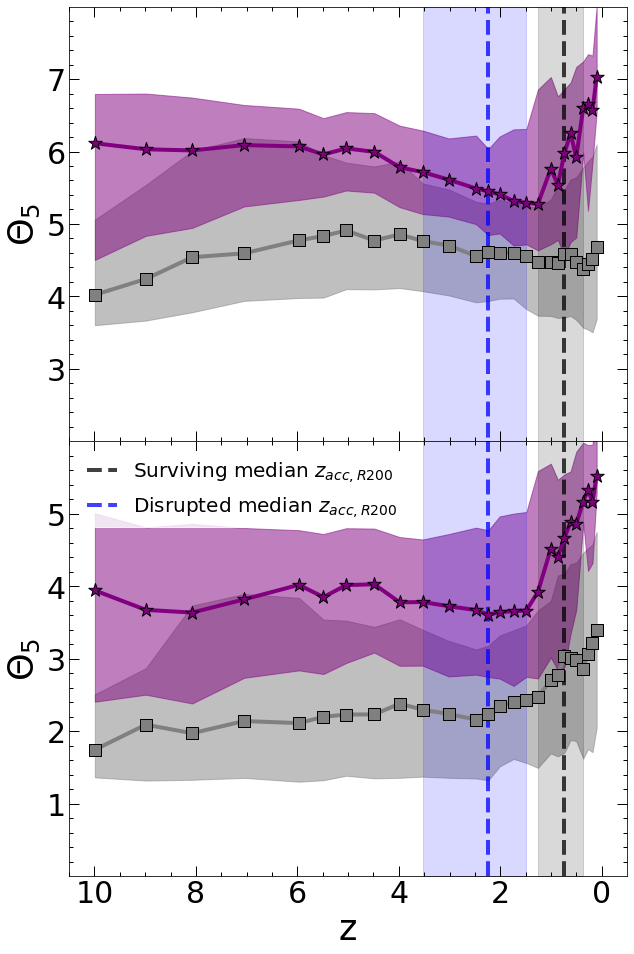}
\caption{{\it Top:} Evolution of the tidal parameter $\Theta_5$ (computed using $M_{n, dyn}$ masses) with $z$, for both the surviving (grey) and disrupted (purple) dwarfs. Solid lines represent the median relations for each population, while the purple and grey shaded regions enclose the corresponding $25^{\rm th}$ and $75^{\rm th}$ percentiles of these medians, respectively. Dashed blue and black vertical lines are the median of all redshifts when the populations cross the virial radii of their hosts (i.e., the median of $z_{\rm{acc}|{\rm R}_{200}}$), while the blue and grey vertical shades represent the $25^{\rm th}$ and $75^{\rm th}$ percentiles around these medians. Higher $\Theta_5$ values for disrupted progenitors across all $z$ compared with $\Theta_5$ values of surviving dwarfs, indicate that the disrupted progenitors are in denser environments. {\it Bottom:} Similar to the top panel, but using $M_{n, bar}$ for computing $\Theta_5$.}
\label{fig:Evo_Theta}
\end{figure}

\begin{figure*}
\centering
\includegraphics[width=1.8\columnwidth]{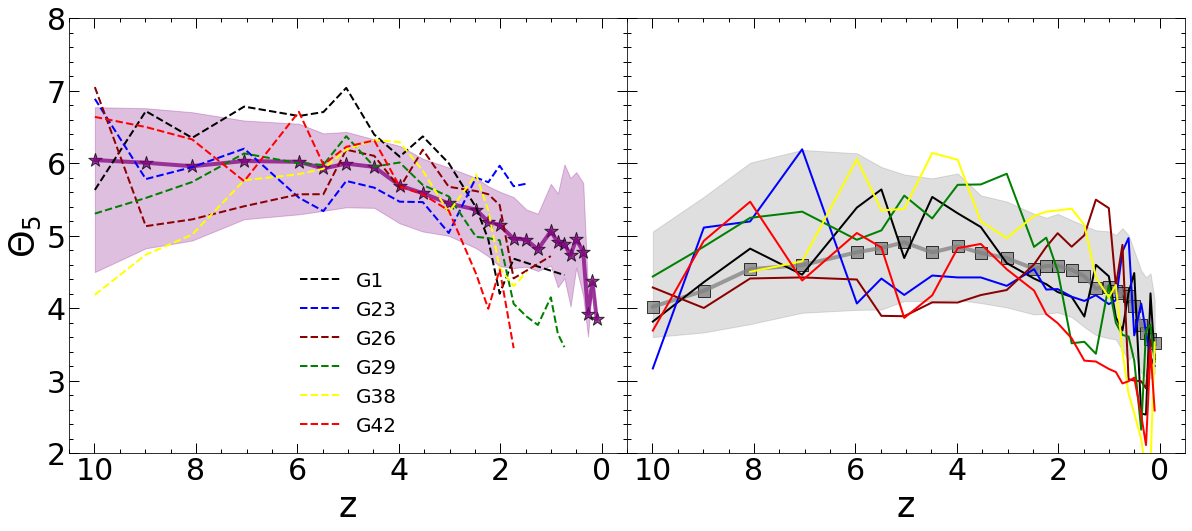}
\caption{Evolution of the median tidal parameter $\Theta_5$ versus redshift, and their associated $25$ and $75$ percentiles around the median for the populations in all simulated hosts (purple line and shaded region for disrupted dwarfs, in the left panel; and grey line and shaded region for the surviving dwarfs, in the right panel). Superimposed are the corresponding $\Theta_5$ medians for the two populations in the $s_6$* subsample. These are indicated with dashed lines for disrupted populations and full lines for the surviving ones. Lines for individual hosts in the  $s_6$* subsample are shown with different colours, as stated in the legend. The evolution of $\Theta_5$ is similar in the case of $s_6$* subsample as in the entire sample.}
\label{fig:Theta_5_six_selected_galaxies}
\end{figure*}

\subsection{The environments of disrupted and surviving dwarfs}
\label{ssec:environment}

The chemical abundances of the dwarf progenitors can also be affected by the environment in which they form and evolve prior to accretion. Here we investigate the scenario proposed by  \citet{Naidu2022}, in which the disrupted population of dwarfs originate in a denser environment, closer to the MW host, than the surviving population. To test this in the simulations, we track the location of the dwarf progenitors at different redshifts and estimate the environments in which they evolve. 
 
To quantify the environment, we use the tidal parameter $\Theta_5$ proposed by \citet{Karachentsev2013}, and defined as:

\begin{equation}
\centering
\Theta_5 = \log \left(\sum_{n=1}^5 \frac{{M}_{n}}{D_{in}^{3}} \right),
\end{equation}

\noindent where ${M}_{n}$ are the masses of the $5$ nearest galaxy neighbours to dwarf galaxy $i$ and $D_{in}$ are the physical distances between these neighbours and the given dwarf galaxy at a given redshift $z$.  This is effectively a measurement of the tidal forces that nearby galaxies exert onto a given dwarf at that redshift. The individual $\Theta_5$ values for each satellite dwarf galaxy/progenitor are computed until the object crosses the virial radius of its host, namely the ${R}_{200}$ radius. 
We also consider two ways to compute the masses of the neighbours: either as the cold baryonic masses, ${M}_{n, bar}={M}_{*} + {M}_{\rm SF}$, or as the total (dynamical) masses, ${M}_{n, dyn} = {M}_{*} + {M}_{\rm SF} + {M}_{\rm NSF} + {M}_{\rm DM},$ where the last two terms indicate the mass of hot or diffuse gas (not star-forming), and of dark matter, respectively. The first method aims to reproduce the one used by \citet{Karachentsev2013} for the observations, where masses are inferred from the luminosities of galaxies, which are then multiplied by a constant factor. The second method is more physically motivated, as it tracks the intrinsic variations in mass-to-light ratios across different masses. Here we consider all galaxies as potential neighbours for the computation of $\Theta_5$, including the ``main haloes'', i.e., the progenitors of the central galaxies corresponding to the MW-mass hosts.

Fig.~\ref{fig:Evo_Theta} shows the evolution of $\Theta_5$ with redshift for both the disrupted and surviving dwarf populations, and also using both methods for computing $\Theta_5$ (using ${M}_{n, dyn}$ in the top panel, and ${M}_{n, bar}$ in the bottom one). The solid lines represent the median $\Theta_5$ values for the disrupted populations (purple) and surviving ones (grey), respectively. 
The dashed vertical lines in this figure are the median redshifts associated with the dwarf satellites crossing for the first time the $R_{200}$ boundaries of their hosts, $z_{\rm{acc}|R_{200}}$. The dashed blue line corresponds to the median $z_{\rm{acc}|R_{200}}$ of the disrupted population, while the dashed black line corresponds to the median of the surviving one. Likewise, the shaded blue and grey vertical shaded regions represent the corresponding $25^{\rm th}$ and $75^{\rm th}$ percentiles around these two medians, respectively. 

This figure shows that the progenitors of disrupted dwarfs have higher $\Theta_5$ values than the surviving satellites, across all redshifts considered here. This is true for both methods of computing $\Theta_5$ (either using $M_{n,dyn}$ or $M_{n,bar}$). Overall, these results suggest that the progenitors of the disrupted dwarfs form and evolve in much denser environments than the surviving dwarfs, which is in agreement with the scenario proposed by \citet{Naidu2022} for the MW populations.  Moreover, our simulations indicate that this behaviour is generally valid for the populations of all MW-mass hosts, not just the MW.

Fig.~\ref{fig:Evo_Theta} also shows that the progenitors of disrupted satellites cross the boundary ${R}_{200}$ of their hosts earlier than the surviving satellites. The corresponding median $z_{\rm{acc}|{\rm R}_{200}}$ redshifts for the disrupted and surviving populations are $\approx 2.2$ and $\approx 0.8$, respectively. Note that these populations comprise satellites from all simulated MW-mass hosts, whereas the individual median $z_{\rm{acc}|{\rm R}_{200}}$ values for populations pertaining to each host may vary considerably (see the vertical shaded regions).  However, for each host, the progenitors of disrupted satellites are accreted earlier than the corresponding surviving population, as expected. 

The median $\Theta_5$ values for the disrupted population show also an interesting variation with redshift, staying relatively constant until $z \approx 4$ and then increasing until close to $z=0$. This is mainly the result of our choice of including the main haloes as neighbours in the calculation of $\Theta_5$. At low redshift the progenitors of both populations are located mainly in the vecinity of their hosts.

Fig.~\ref{fig:Evo_Theta} shows consistent patterns in the types of environments in which disrupted and surviving populations evolve, across all simulated MW-mass systems. However, there is a significant amount of scatter in the $\Theta_5$ values at any given redshift. In Fig.~\ref{fig:Theta_5_six_selected_galaxies} we investigate in more detail the evolution of the $\Theta_5$ parameter for the two populations in the six simulated MW-mass systems which are more similar to the MW, namely the $s_6$* subsample. This allows us to assess the variations in $\Theta_5$ versus $z$ on a system by system basis, and also explore any differences that may occur in MW-like hosts. This figure shows that the median $\Theta_5$  values for the disrupted dwarfs of these systems are roughly within the $25^{\rm{th}}$ and $75^{\rm{th}}$ percentiles around the medians of the corresponding simulated samples including all hosts, although, occasionally, the $\Theta_5$ values reach above and below these limits. Note also that the $\Theta_5$ parameters are not computed up to $z=0$ in the case of the disrupted dwarfs in these systems, given that these populations are typically accreted at redshifts of $\approx 1.5 -2$ and tend to get fully disrupted before present time.

Overall, our findings support the scenario in which the disrupted dwarfs inhabit denser environments than surviving ones. Moreover, we find that this is the case at all redshifts, and that is a common pattern in all MW-mass hosts, regardless of whether they have similar accretion histories to the MW (i.e., $s_6$* subsample) or not.

\begin{figure*}
\centering
\includegraphics[width=1.6\columnwidth]{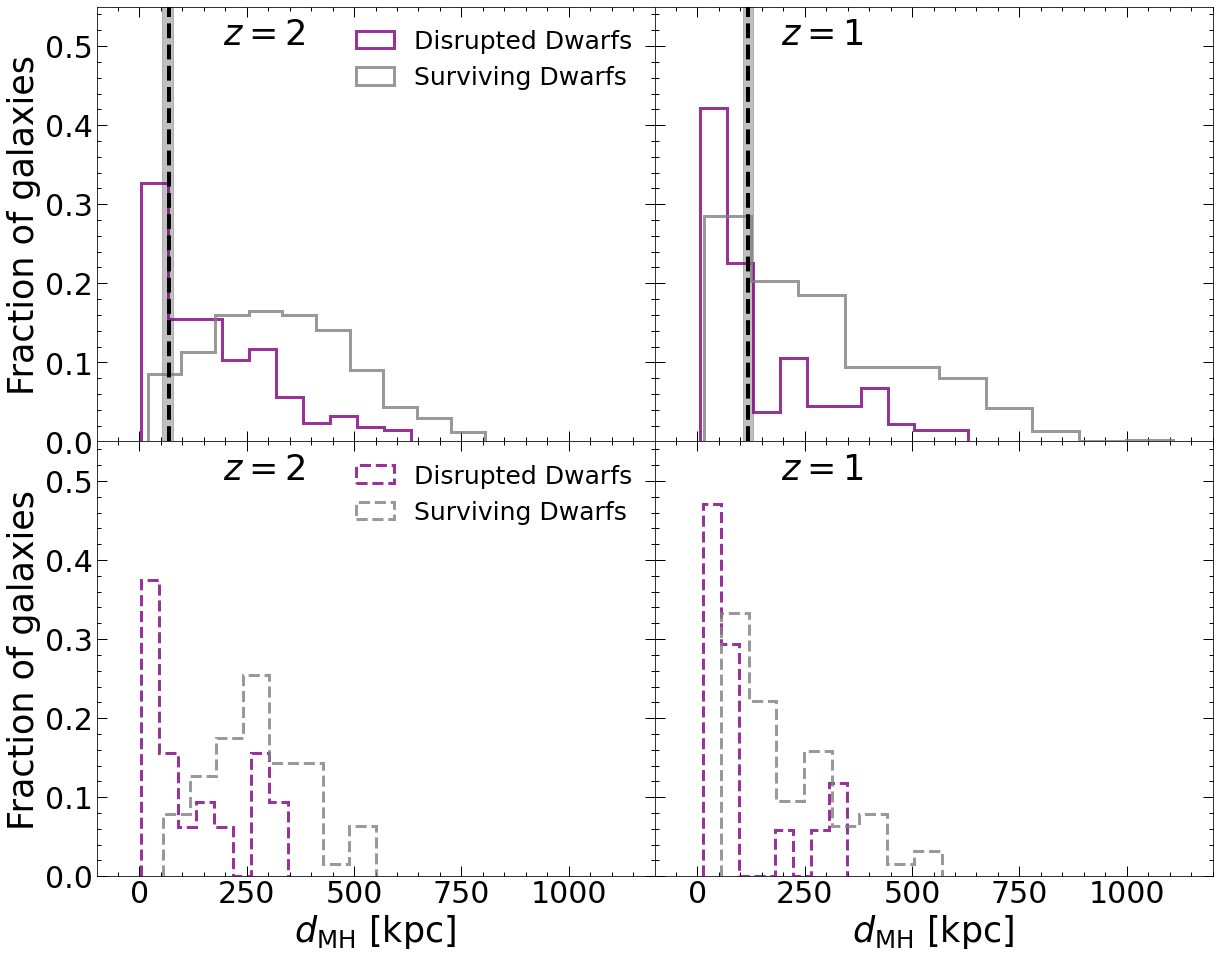}
\caption{{\it Top:} Distributions of distances between the progenitors of dwarf galaxies (disrupted dwarfs with purple lines and surviving ones with grey lines, respectively) and the centers of potential of the corresponding progenitors of their MW-mass hosts, at $z=2$ ({\it left}) and at $z=1$ ({\it right}), respectively. 
The black dashed lines represent the median $R_{200}$ of the progenitors of MW-mass systems at the corresponding redshift, while the shaded regions are the $14^{\rm th}$ and $86^{\rm th}$ percentiles of these values. Dwarf satellites which are disrupted today were closer to their hosts than the surviving ones, both at $z=1$ and at $z=2$. {\it Bottom:} The same distributions as in the top panels, but for the populations in the six hosts in the $s_6$* subsample. The differences between the locations of the two satellite populations are similar to those in the full sample, being slightly more pronounced in this case.}
\label{fig:Histograms-distance}
\end{figure*}

The scenario proposed by \citet{Naidu2022} envisages that the reason that environments of disrupted dwarfs are denser than those of surviving ones, is because the progenitors of the disrupted dwarfs form earlier, and closer to their hosts. To verify whether the simulated populations are located at different distances from their hosts, we trace their evolutionary paths from their time of formation until they the last time they are identified in the simulations.  The top panels in Fig.~\ref{fig:Histograms-distance} show the distribution of physical distances $d_{\rm MH}$ of the progenitors of the two populations relative to the centers of potential of the corresponding progenitors of their hosts (i.e., the main haloes). The $d_{\rm MH}$ distances are shown at two fixed redshifts, $z=2$ and at $z=1$ (left versus right panels), for both disrupted and surviving satellites (purple and grey histograms, respectively). 

In general, the progenitors of disrupted dwarfs are located significantly closer to their hosts than the surviving dwarfs, at both redshifts. For example, at $z=2$, about a third of the progenitors of disrupted dwarfs are already within the virial radius of their host (shown with vertical dashed line in this figure) whereas less than $10\%$ of surviving dwarfs are that close. By $z=1$, about $60\%$ of disrupted progenitors are within the virial radius of their hosts, compared with less than $30\%$ of surviving ones. The shapes of the distributions are also different, the distributions for disrupted satellites being narrower than those of the surviving ones, at both redshifts. The differences are more well defined at $z=2$ than at $z=1$, suggesting that these differences are generated early on, likely from the times of formation of these populations.

The bottom panels of Fig.~\ref{fig:Histograms-distance} show the same analysis, but for the populations in the $s_6$* subsample. Here each population (disrupted / surviving) combines all dwarf systems of this type from the six MW-mass hosts in the subsample. 
The results are similar to the case of the entire sample, namely that the progenitors of the disrupted dwarfs are closer to their hosts than the surviving population, at both redshifts. The differences appear to be slightly stronger in this case (however, note the smaller sample size). Specifically, at $z=2$, the  distribution peaks at distance below $100$~kpc from the host (which is well within the $R_{200}$ at that time), whereas the corresponding distribution for surviving satellites peaks around $250$~kpc (i.e., outside that radius). By $z=1$, $\simeq 78\%$ of progenitors of disrupted dwarfs are within $d_{\rm MH} = 75$~kpc from their hosts, compared with $\sim 66$ of surviving satellites.

In summary, the trends seen in Figs.~\ref{fig:Evo_Theta}, ~\ref{fig:Theta_5_six_selected_galaxies} and ~\ref{fig:Histograms-distance} are consistent with the scenario proposed by \citet{Naidu2022}, in which disrupted dwarfs form in denser environments, closer to their hosts. We also show that the behaviour is common for all MW-mass hosts, not just the ones matching the MW's accretion history. Additionally, we have shown that the progenitors of disrupted dwarfs continue to evolve in denser environments than the surviving satellites, although the differences between the two populations tend to decrease towards low redshift (below 1).  

\begin{figure*}
\includegraphics[width=1.8\columnwidth]{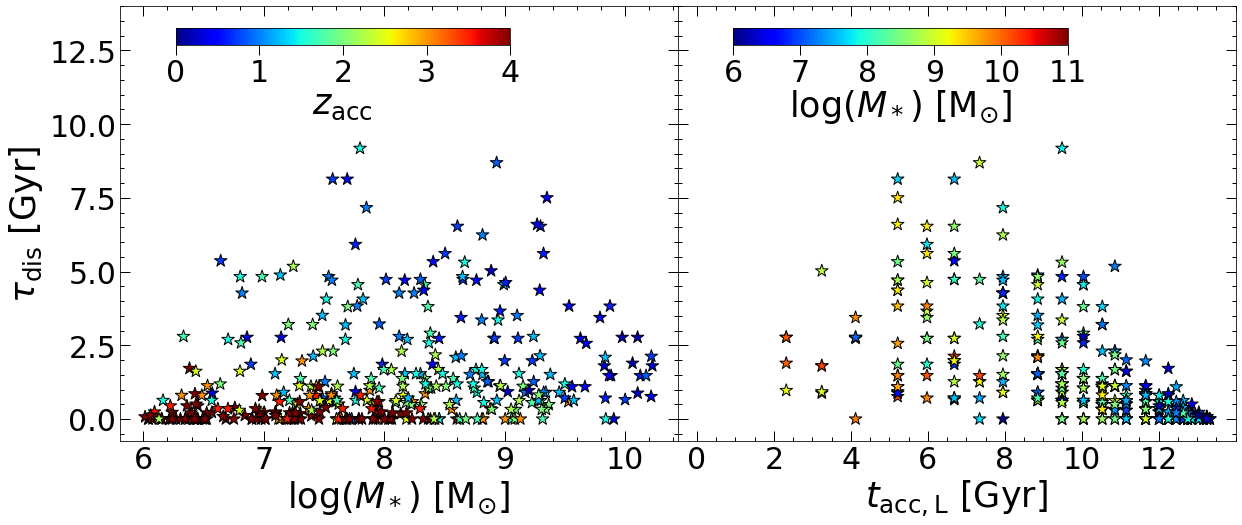}
\caption{{\it{Left:}} Disruption time, $\tau_{\rm{dis}}$, versus ${M}_*$ for disrupted dwarfs, with symbols colour-coded according to $z_{\rm{acc}}$. The most recently accreted dwarfs tend to have higher masses and greater $\tau_{\rm{dis}}$ values. {\it{Right:}} $\tau_{\rm{dis}}$ versus lookback time of accretion, $t_{\rm{acc,L}}$, for the disrupted dwarfs. Values are colour-coded by ${M}_*$.}
\label{fig:tau_dis_colormap_de_z_acc}
\end{figure*} 

\subsection{Disruption times}
\label{ssec:disruption}

For the disrupted population we have so far focused mainly on their properties at the time of accretion ($z_{\rm acc}$), or prior to that.  In this section, we estimate the timescales in which they disrupt after they are accreted onto their hosts. As mentioned earlier, if dwarf galaxies spend long periods of time orbiting inside their hosts they may experience additional star formation and therefore more chemical evolution. We expect that mostly high mass satellites could be undergo additional star formation, as low mass satellites are effectively stripped of their cold gas and quenched at (or even before) accretion, as discussed in Section~\ref{ssec:scatter}. 

We therefore estimate the disruption times for the progenitors of satellites fully disrupted by $z=0$, in all MW-mass hosts. These are computed as the time intervals $\tau_{dis}$ between the first time when the progenitors cross the virial radii (${\rm R}_{200}$) of their hosts, $t_{\rm{acc}|{\rm R}_{200}}$, and the last time\footnote{Given the limited number of simulation outputs, both, $t_{\rm{acc}|R_{200}}$ and $t_{\rm f}$, correspond to approximated measurements of the actual values. In practice however, we consider $t_{\rm{acc}|{\rm R}_{200}}$ as the time associated with the first snapshot when \texttt{SUBFIND} identifies the dwarf as a subhalo inside the ${R}_{200}$ boundary of the host, while $t_{\rm f}$ corresponds to the last snapshot when \texttt{SUBFIND} identifies the dwarf as a gravitationally bound subhalo.} when they are identified as individual subhaloes by \texttt{SUBFIND} in the simulations, $t_{\rm f}$. Therefore,

\begin{equation}
\tau_{\rm{dis}} = t_{\rm f} - t_{\rm{acc}|{\rm R}_{200}}.    
\end{equation}  

\noindent The redshift associated with the progenitor crossing the ${\rm R}_{200}$ boundary of its host has been introduced before, and denoted $z_{\rm{acc}|{\rm R}_{200}}$. Note that we take into account here only satellites which are fully disrupted at $z=0$. According to our definition, those satellites which are still in the process of disruption at $z=0$ are considered as surviving.

The left panel of Fig.~\ref{fig:tau_dis_colormap_de_z_acc} shows the distribution $\tau_{\rm{dis}}$ versus the maximum stellar masses of the corresponding progenitors (this is $M_*$ measured at $z_{\rm acc}$, as discussed in Section~\ref{sec:merger_trees}). These are shown as star symbols, which are colour-coded by $z_{\rm acc}$ of each progenitor. This figure shows that progenitors which are accreted very early (e.g., $z_{\rm acc}$ between $3-4$) are typically of low mass ($M_* \lesssim 10^8\, {\rm M}_{\odot}$), and have very short disruption times ($\tau_{dis} \lesssim 1$~Gyr). This is expected given that the progenitors of the  hosts are small in size at high redshift, and therefore the timescales on which satellites merge with the main halo at these epochs are very short. Satellites that infall at $z_{\rm acc }\lesssim2$ (green and blue tones) typically have higher masses, which is also expected as they have more time to form stars prior to accretion. As their hosts are also larger in size at low redshift, satellites tend to take longer to disrupt. This figure shows that massive satellites can have a wide range of disruption times (typically, between $0-6$~Gyr for those accreted at $z_{\rm acc} \simeq 2$ and up to $9-10$~Gyr for those accreted  $z_{\rm acc} \lesssim 1$). The longer disruption times for more massive satellites, coupled with them being more gas-poor at accretion (see Figs.~\ref{fig:MZR-z50-fSF} and \ref{fig:alpha-Mstar-z50-fSF}), suggests the possibility of some of them undergoing additional star formation inside their hosts.
However, this effect is probably not significant, because, as suggested by Fig.~\ref{fig:zacc_vszacc200_Mstar}, most massive progenitors have already reached their maximum stellar mass by the time of accretion. 

The right panel of Fig.~\ref{fig:tau_dis_colormap_de_z_acc} shows a similar analysis, this time plotting $\tau_{\rm{dis}}$ versus the lookback time of accretion, $t_{\rm{acc,L}}$. The latter computes how long ago the progenitors have been accreted onto their host (compared to present-day), whereas $\tau_{dis}$ measures the time taken to disrupt since accretion time $t_{\rm{acc}|{\rm R}_{200}}$. The star symbols here are colour-coded by stellar mass at accretion, ${M}_*$. This panel shows a similar result as the left panel, in that progenitors accreted a long time ago ($t_{\rm{acc,L}} \simeq 10-12$~Gyr) are typically low mass ($M_* \simeq 10^6 - 10^7\, {\rm M}_{\odot}$, blue tones) and are disrupted quickly (in less than $1$~Gyr). Progenitors accreted between $6-9$~Gyr ago show the largest range in $\tau_{dis}$. Interestingly, the scatter in $\tau_{dis}$ does not appear to depend on the masses ($M_*$) of progenitors, suggesting that other factors are at play (e.g., the total mass of the host, or the orbital properties of the infalling dwarfs). Satellites accreted within the past $2-5$~Gyr have short disruption times (between $0-5$~Gyr) and are fairly massive. However the short disruption times are mainly a selection effect, as most recently accreted satellites are not yet fully disrupted, and therefore are not included in this plot. Also, no satellites accreted less than $2$~Gyr ago are fully disrupted in our simulated sample, for the same reason.

\section{Discussion: The evolution of the MZR with redshift}
\label{sec:discussion}

Our results give insight into the stellar mass -- (stellar) metallicity relation of the disrupted population of satellite dwarf galaxies in MW-mass systems. The simulations indicate how this relation evolves with redshift (Fig.~\ref{fig:Disrupted-MZR-different-z-single-panel}) and predict correlations of the scatter in [Fe/H] at fixed $M_*$ with other characteristics of the accreted dwarfs, namely with their redshift of accretion ($z_{\rm acc}$) and the cold gas fraction ($f_{\rm SF}$) at the time of accretion (Fig.~\ref{fig:MZR-z50-fSF}). Here we discuss these findings in the context of previous work.

The stellar mass -- stellar metallicity relation has been studied extensively in the Local Universe, where observations show it is a global scaling relation for galaxies which extends over many orders of magnitude in stellar mass  \citep[e.g.,][]{Gallazzi2005,Gallazzi2014,Trussler2020}. This scaling relation continues at the low-mass end of galaxy formation, in the regime of ``classical'' and even ultra-faint dwarf satellites of the MW \citep{Kirby2013,Simon2019}. Cosmological simulations qualitatively match this relation over a wide range of stellar masses \citep[e.g.,][]{Schaye2015,Ma2016}. A similar scaling relation is observed between the stellar masses and the gas-phase metallicities of galaxies \citep[e.g.,][]{Lequeux1979,Skillman1989,Tremonti2004,Kewley2008,Maiolino2008, Sanders2021}, which is also broadly matched by the simulations  \citep[e.g.,][]{DeRossi2015,Ma2016,Dave2017,DeRossi2017,Torrey2019}.

Several models have been proposed to explain the origin and shape of the MZR. These include: strong stellar feedback \citep{Dekel1986}, where supernovae winds expel the metal-enriched gas from low-mass galaxies due to their shallower potential wells; an equilibrium metallicity in galaxies \citep{Finlator2008}, which is obtained by balancing the gas inflow rate with the gas consumption rate from star formation plus the gas outflow rate;by energy-driven supernova winds \citep{Guo2016,Dave2012}, by inefficient star formation in dwarf galaxies \citep{Brooks2007}, or by a top-heavy initial mass function \citep{Koppen2007}.

The validity of these models can be tested by observing  the evolution of the MZR with redshift and comparing it with the model predictions. However, at higher $z$, the MZR is much less well constrained, in both observations and simulations. While it is well established that, at fixed ${M}_*$, the high $z$ galaxies have lower metallicities than those in the Local Universe (e.g., \citealt{Erb2006,Sommariva2012,Gallazzi2014, Leethochawalit2018}), there are currently uncertainties in terms of the normalisation of the relation \citep{Kewley2008} or the rate of increase in metallicity at different redshifts  \citep{Maiolino2019}. 

Some comparisons of our results (Fig.~\ref{fig:Disrupted-MZR-different-z-single-panel}) with the observations are possible, particularly at the higher mass end of the simulated dwarfs. For systems in the range of $8.5 < \log({\rm M}_{*} / {\rm M}_{\sun}) < 10.2$, we find a good agreement with the data from the VANDELS survey \citep{Cullen2019}, where a metallicity offset of $\approx 0.6$~dex is found between local galaxies and those at $2.5 < z < 5.0$. Our predicted MZR evolution at the high-mass end of dwarf satellites is also similar to that obtained by \cite{Gallazzi2014} for their sample of massive galaxies ($M_* > 10^{10}~{\rm M}_{\sun}$) at $z\approx 0-0.7$.  

Overall, we find a gradual decrease of the normalization of the MZR with redshift (see Fig.~\ref{fig:Disrupted-MZR-different-z-single-panel}). This behaviour is consistent with the evolution of the MZR obtained in other simulations, for example, in the IllustrisTNG \citep{Torrey2019} 
The driving force behind the evolution of the MZR is likely the fraction of cold gas in galaxies. This would also explain the correlation of the scatter in the MZR with the gas fractions at different redshifts (see Figs.~\ref{fig:MZR-scatter-fits} and~\ref{fig:disrupted-fSFvsalpha-cmap}), which is seen in other cosmological simulations \citep[e.g.,][]{Ma2016,Dave2017,DeRossi2017,Torrey2019}. The main trend obtained in our study, namely that more recently accreted dwarfs have lower gas fractions and higher metallicities, is in general agreement with previous results. Another physical process that might contribute to the evolution of the MZR is the increasing occurrence of Type Ia supernova as redshift decreases, which raises the concentration of iron in the gas that will form new stars (e.g., \citealt{Leethochawalit2019}).

We note that the Auriga simulations also show differences in [Fe/H] between the surviving and disrupted satellites at fixed stellar mass, and that the scatter in [Fe/H] of surviving satellites depends on the infall time, at fixed $M_*$  (see figure 5 in \citealt{Fattahi2020}). However, the scatter in [Fe/H] of disrupted satellites does not show a clear dependence with infall time (Fattahi, priv. comm.). One reason for this could be due to slightly different definitions of infall and accretion times. A more detailed comparison between the two simulations may shed light on these differences, however we leave this for a future study.

The evolution of the slope of the MZR is another constraint on the models. Our results (Fig.~\ref{fig:Disrupted-MZR-different-z-single-panel}) suggest that the slope does not change significantly with redshift, at least in the regimes investigated here, of dwarf satellites and up to $z \approx 4$. This behaviour is also in agreement with observations \citep[e.g.,][]{Zahid2014,Sanders2021}. 

Our simulations also indicate a steeper MZR slope for the combined population of disrupted satellite galaxies (of $\approx 0.48$) than for the MZRs of satellites accreted at any given redshift (slopes of $\approx 0.3$). This prediction can be further tested with improved observations in the MW, in particular by constraining the properties of the lowest mass/ earliest accreted satellite galaxies, such as I'itoi or other similar systems.

\section{Conclusions}
\label{sec:concl}

We used the \texttt{ARTEMIS} suite of zoomed-in, cosmological hydrodynamical simulations to investigate the properties of the satellite dwarf galaxies in MW-mass hosts. The simulations comprise 42 MW-mass hosts, with a variety of accretion histories. We focused on comparing the properties of disrupted satellites versus those surviving at $z=0$, and on comparisons with observations of these populations in the MW. We found several properties of disrupted satellites that connect with their redshift of accretion, and therefore can be used to constrain the accretion history of MW-type galaxies. Our main results can be summarised as follows:

\begin{itemize}

\item Disrupted dwarf galaxies follow a stellar MZRs (Fig.~\ref{fig:MZR-scatter-fits}), but with lower normalisation than the MZR for the surviving dwarfs (lower [Fe/H] at fixed  $M_*$). Combining progenitors of disrupted dwarfs from all MW-mass hosts, and accreted at all redshifts, we find an average slope of the MZR of $\approx0.48$, which is steeper than the slope estimated for the disrupted populations of the MW \citep{Naidu2022}, of $\approx 0.3$. 
On an individual basis, however, the MZR slopes of the disrupted dwarfs in each MW-mass system can vary significantly, and some of them can match the MW observations. We find better agreement with the observational data if Sagittarius and I'itoi are included in the observational sample (in this case the observational MZR would have a slope of $\approx 0.5)$. For the surviving populations, we find an excellent agreement between the predicted MZR slope, of $\approx 0.32$, and the sloped measured from observations in the MW and Local Group \citep{Kirby2013,Naidu2022}, of $0.3$.

\item  We investigated the evolution of the MZR of the disrupted population with redshift, and found that populations accreted at the similar redshifts ($z_{\rm acc}$) follow MZRs with similar slopes as the surviving populations, of $\approx 0.3$, and with lower normalizations towards higher redshift   (Fig~\ref{fig:Disrupted-MZR-different-z-single-panel}).
The steeper slope obtained for the combined  population of disrupted dwarfs ($\approx0.48$), compared with the slope derived for these populations at fixed redshift (of $\approx 0.3)$, can be explained by the changes in the spectrum of stellar masses of dwarfs accreted towards lower redshift, and their ongoing metal-enrichment. For disrupted dwarfs accreted most recently ($z_{\rm acc} \approx 0.5$), their MZR is similar to that of surviving population.

\item Disrupted dwarf satellites have significantly different chemical properties than the surviving ones:  at fixed $M_*$, they have lower [Fe/H] (Fig.~\ref{fig:MZR-scatter-fits}) and higher [Mg/Fe]  (Fig.~\ref{fig:alpha-vs-Mstar}); also, they have higher [Mg/Fe], at fixed [Fe/H]. These trends are consistent with observations in the MW.

\item The [Fe/H] -- $M_*$ and [Mg/Fe] -- $M_*$ distributions of disrupted dwarfs show a significant amount of scatter. We find that this scatter correlates with both $z_{\rm acc}$ and the cold gas fractions at accretion, $f_{\rm SF}$ (Figs.~\ref{fig:MZR-z50-fSF} and~\ref{fig:alpha-Mstar-z50-fSF}), of the progenitors of these dwarfs. Specifically, at fixed $M_*$, progenitors accreted earlier have lower [Fe/H], higher [Mg/Fe] and also higher $f_{\rm SF}$ than those accreted more recently.  

\item Generally, the cold gas fraction of the disrupted dwarfs correlates with their $z_{\rm acc}$, but not always. Typically, before  $z_{\rm acc} \approx 4$, dwarf galaxies are gas rich (with $f_{\rm SF}$ roughly constant, of $\approx 80-100\%$). Below $z\approx 4$, however, $f_{\rm SF}$ decreases sharply (Fig.~\ref{fig:disrupted-fSFvsalpha-cmap}). A small fraction of the disrupted dwarfs in the simulations do not follow these general trends. These are low-mass systems with $f_{\rm SF}=0$, and are likely an artifact of the limited numerical resolution.

\item We investigated several global properties of the disrupted populations to test how predictive they are in terms of their host accretion history. Among these properties, we find that the mass-weighted values $\langle {\rm [Fe/H]} \rangle$ and $\langle {\rm [Mg/Fe]} \rangle$ of each disrupted population correlate strongly with the corresponding $\langle z_{\rm acc} \rangle$ value. In contrast, the average slopes of the MZR or of the [Mg/Fe]--$M_*$ relation in the disrupted population are not good tracers of the accretion history. Furthermore, we selected the six disrupted populations (the $s_6$* subsample) with similar $\langle {\rm [Fe/H]} \rangle$ and $\langle {\rm [Mg/Fe]} \rangle$ values as the ones in thee disrupted population in the MW (Fig.~\ref{fig:Weighted_alpha_vs_Z_zacc_cmap}), and found that these were accreted, on average, at $\langle z_{\rm acc} \rangle \simeq 1.8$, which is in agreement with results from observations (Fig.~\ref{fig:Sl.opes_and_Z_vs_weighted_zacc}).

\item The disrupted dwarfs differ from the surviving counterparts also in terms of their stellar mass function.  For surviving dwarfs, there are typically more low-mass dwarfs than high mass, as expected. However, for the combined disrupted population (i.e., accreted at all $z_{\rm acc}$), the distribution is more evenly spread. This supports the differences seen in the global MZRs and other distributions between the two populations. There is also an indication that galaxies with chemical properties like the MW (i.e., the $s_6$* subsample) may have a higher fraction of low-mass dwarfs in their disrupted populations than other galaxies of MW-mass (Fig.~\ref{fig:Disrupted_mass_distribution}).

\item Simulations also support the scenario in which the disrupted dwarfs of the MW form and evolve in denser environments than the surviving ones (Figs.~\ref{fig:Evo_Theta}). This is because the progenitors of disrupted dwarfs form closer to their hosts (Fig.~\ref{fig:Histograms-distance}). Differences in the environments and locations of the progenitors of the two populations are more predominant at high redshift (e.g. $z=2$) than at lower one ($z=1)$. These differences are found in all MW-mass simulated galaxies.

\end{itemize}

\section*{Acknowledgements}
The authors thank the referee for useful suggestions which helped improved our paper, and acknowledge discussions with Azi Fattahi about the Auriga simulations. The authors thank the members of the \texttt{EAGLE} team for making their cosmological simulation code available for the \texttt{ARTEMIS} project and John Helly for constructing the merger trees for the MW-mass haloes studied here. We used various Python libraries, including \textsc{H5Py} (\url{http://www.h5py.org/}) and \textsc{AstroPy} \citep{Astropy2022}, as well as the publicly available \textsc{read\_eagle} module (\url{https://github.com/jchelly/read_eagle}, \citealp{eagle2017}). The {\texttt ARTEMIS} project has received funding from the European Research Council (ERC) under the European Union's Horizon 2020 research and innovation programme (grant agreement No 769130).  We acknowledge support from {\it Agencia Nacional de Promoci\'on de la Investigaci\'on, el Desarrollo Tecnol\'ogico y la Innovaci\'on} (Agencia I+D+i, PICT-2021-GRF-TI-00290, Argentina). 

\section*{Data availability}
 Data from \texttt{ARTEMIS} simulations or merger trees may be shared on reasonable request to the corresponding author.


\bibliographystyle{mnras}
\bibliography{refs} 




\appendix

\section{Properties of the disrupted and surviving populations in the MW-type hosts}
\label{sec:s_6}

Here, we investigate in more detail the MZR and the [Mg/Fe] -- ${\rm M}_{*}$ relations for the disrupted and surviving dwarf populations in the $s_6$* subsample (i.e., in the six MW-mass hosts indicated with green circles in Fig.~\ref{fig:Sl.opes_and_Z_vs_weighted_zacc}; see also the discussion in Section~\ref{ssec:av_prop}). 

\begin{figure*}
\centering
\includegraphics[width=1.6\columnwidth]{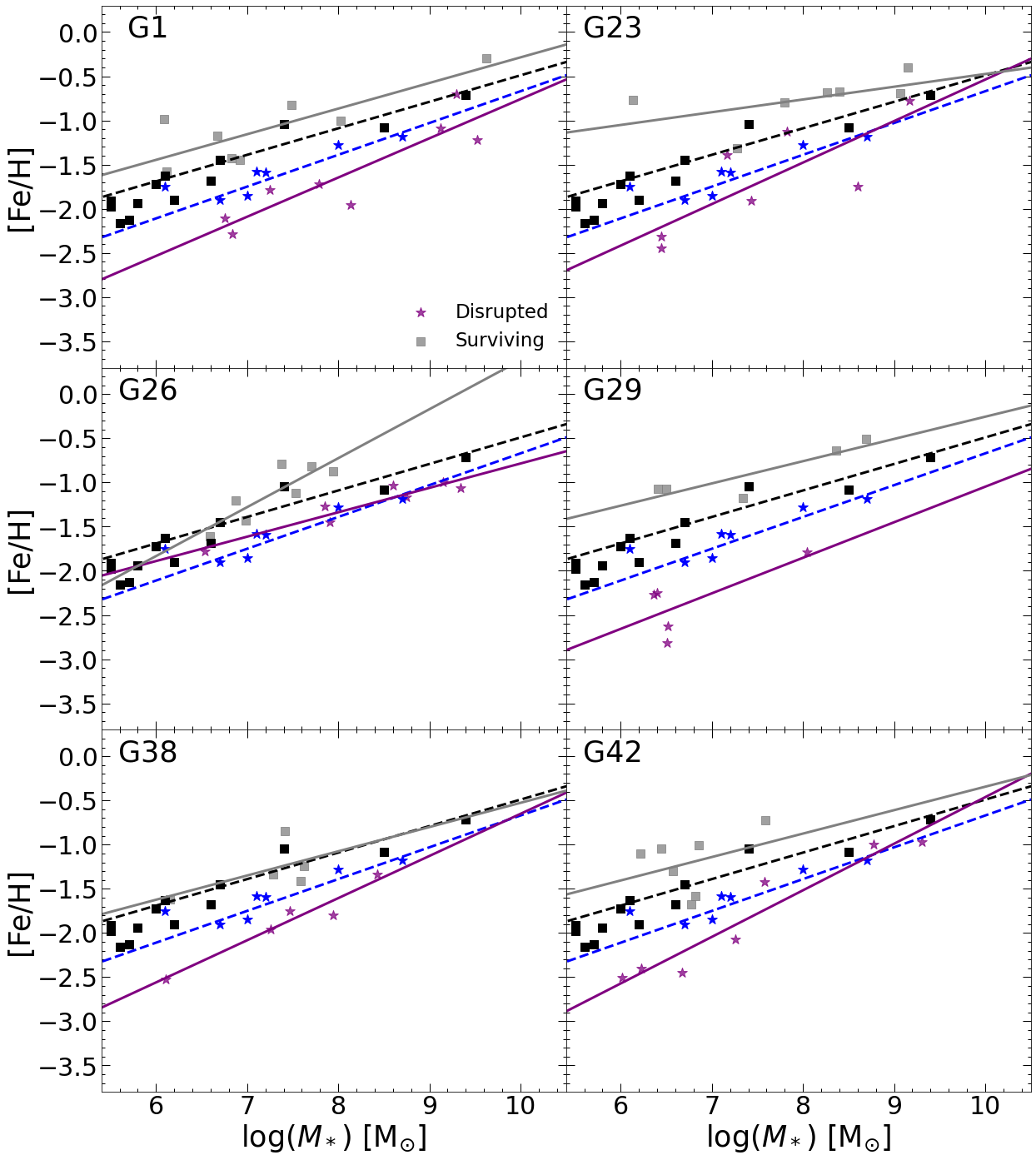}
\caption{MZRs for the disrupted (purple stars) and surviving (grey squares) populations in the $s_6$* hosts, and the corresponding MW populations studied by \citet{Naidu2022}  (blue stars and black squares, respectively). The linear fits for the simulated dwarfs are shown with purple and grey solid lines, respectively. The dashed blue and black lines correspond to  similar fits to the MW dwarf populations, where the disrupted sample excludes I'itoi and Sagittarius.}
\label{fig:6-galaxies-MZR}
\end{figure*}

The MZRs of both the surviving and disrupted dwarf populations, and their best linear fits,  
are plotted in Fig.~\ref{fig:6-galaxies-MZR}, where each panel corresponds to populations in an individual MW-mass host from the $s_6$* subsample (galaxies G1, G23, G26, G29, G38 and G42). For comparison, we also show the corresponding data for the two populations in the MW. The main properties of these  populations, and the results of the fits, are summarised in Table~\ref{tab:MZR_fits}.

In general, we find that the slopes of the best fits to the simulated MZRs are in agreement with the corresponding slopes associated to the MW populations (for both disrupted and surviving). Specifically, for the disrupted populations, the MZR slopes range from $0.28$ (G26) to $0.53$ (G42), whereas the observed population (MW* sample) has a slope of $0.36$. This agreement is encouraging given that these six populations have been selected based on their averaged chemical abundances ($\langle $[Fe/H]$\rangle$ and $\langle$[Mg/Fe]$\rangle$). For the surviving populations, 4 out of 6 systems have similar MZR slopes (values of $0.25-0.29$) with the surviving population in the MW and the Local Group (slope of $0.3$).

\begin{figure*}
\centering
\includegraphics[width=1.6\columnwidth]{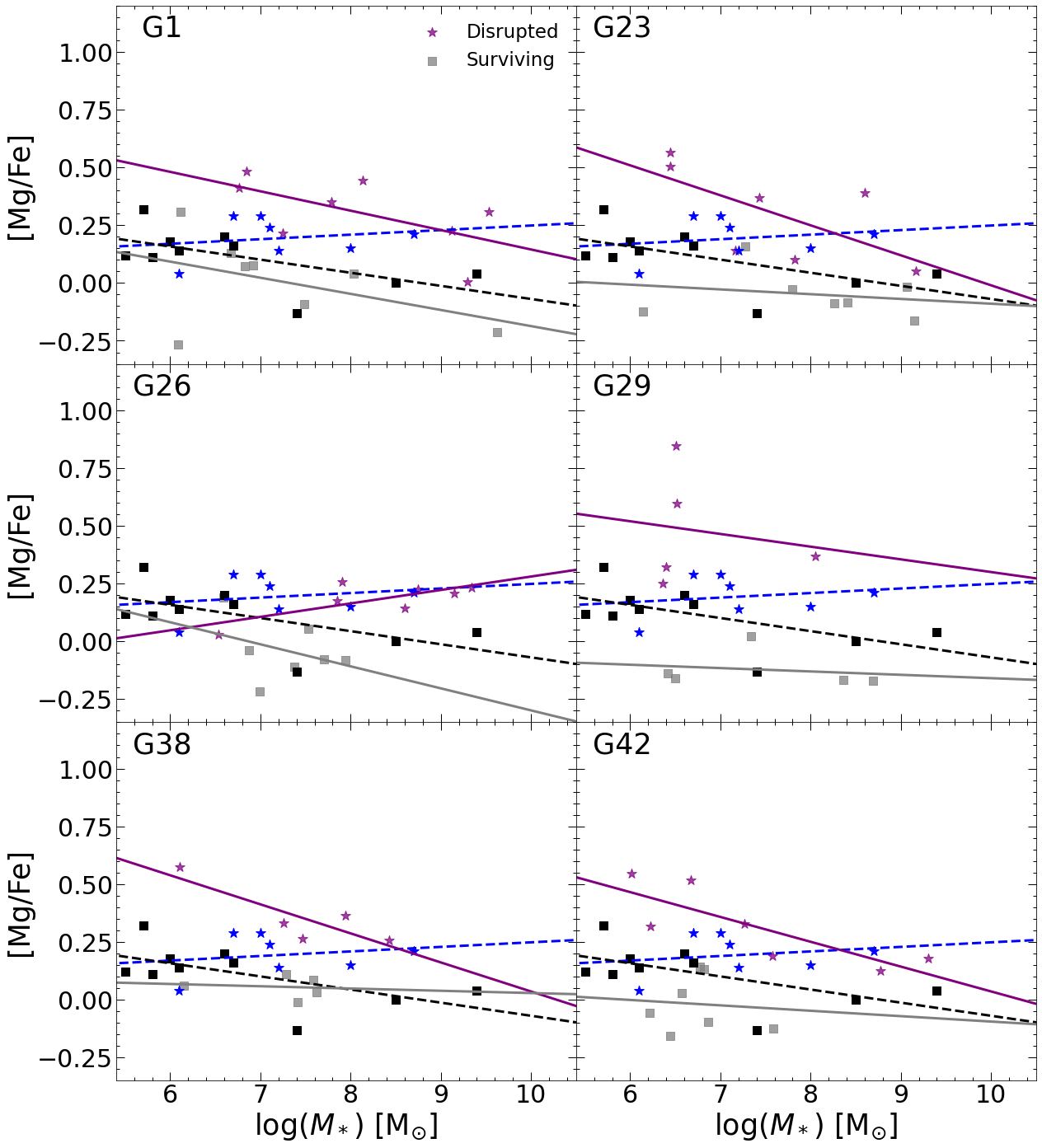}
 \caption{[Mg/Fe]--$M_*$ relation for the disrupted (purple stars) and surviving (grey squares) populations in the $s_6$* hosts and the corresponding MW populations studied by \citet{Naidu2022}  (shown with blue stars and black squares, respectively). The linear fits for the simulated disrupted and surviving populations are shown with purple and grey solid lines, respectively. The dashed blue and black lines correspond to similar fits to the MW dwarf populations. For the observational disrupted sample we exclude I'itoi and Sagittarius (i.e., the MW* sample).}
\label{fig:6-galaxies-alpha-Mstar}
\end{figure*}

The [Mg/Fe] -- ${M}_*$ distributions for the two populations in the $s_6$* hosts, in comparison with the same distributions for the MW populations, are shown in Fig.~\ref{fig:6-galaxies-alpha-Mstar}. The best linear fits to these relations are also included in this figure, while the parameters of these fits are summarized in Table~\ref{tab:alpha_vs_Mstar_fits}. 

In this case, the parameters of the fits (slope and intercept) vary to some extent, and the agreement with observations is not as good as for the MZRs. Such discrepancies could arise because, along our considered ${\rm M}_*$ interval, [Mg/Fe] covers a much smaller range ($\lesssim 0.75$~dex) than [Fe/H] ($\lesssim 3.0$~dex). Hence, the obtained [Mg/Fe] -- ${\rm M}_*$ relations are flatter than the MZRs, suggesting that a linear model might not describe well the behaviour of the former relations, at least for our considered galaxy samples.

\begin{table*}
	\centering
	\begin{tabular}{|c|c|c|}
		\hline
		 & Disrupted & Surviving \\
		\hline
            
		\begin{tabular}{p{1cm}|p{1cm}|p{1cm}}
                Host & $ \langle z_{\rm{acc}} \rangle$ & $ \langle \rm{[Fe/H]} \rangle$\\
                \hline
                MW & 1.20 & -1.09 \\
                MW$^{*}$ & 1.80 & -1.22 \\
                G1 & 2.42 & -1.27 \\
                G23 & 1.76 & -1.21 \\
                G26 & 1.93 & -1.26 \\
                G29 & 2.11 & -1.33 \\
                G38 & 1.90 & -1.34 \\
                G42 & 1.98 & -1.19 \\
                \end{tabular} &
                \begin{tabular}{|p{1.0cm}|p{1.0cm}|p{1.0cm}|}
                Slope & Offset & $r_s$ \\
                \hline
            $0.50$ & $-5.30$ &  \\
            $0.36$ & $-4.27$ &  \\
                $0.44$ & $-5.20$ & $0.83$ \\
                $0.47$ & $-5.24$ & $0.82$ \\
                $0.28$ & $-3.55$ & $0.79$ \\
                $0.40$ & $-5.07$ & $0.20$ \\
                $0.48$ & $-5.42$ & $0.90$ \\
                $0.53$ & $-5.74$ & $0.96$ \\
                \end{tabular}  & 
                \begin{tabular}{|p{1cm}|p{1cm}|p{1cm}|}
                Slope & Offset & $r_s$ \\
                \hline
                $0.30$ & -3.49& \\
                $0.30$ & -3.49& \\
                $0.29$ & $-3.19$ & 0.50 \\
                $0.14$ & $-1.92$ & 0.79 \\
                $0.55$ & $-5.16$ & 0.71 \\
                $0.25$ & $-2.77$ & 0.60 \\
                $0.27$ & $-3.27$ & 0.50 \\
                $0.27$ & $-3.00$ & 0.36 \\
                \end{tabular}  \\
        
		\hline
        
	\end{tabular}
        \caption{Main properties associated to the MW and to the six MW-mass hosts in the $s_6$* subsample shown in Fig.~\ref{fig:6-galaxies-MZR}): the slopes and offsets derived from the fits to the MZR of the disrupted and surviving populations, and their corresponding Spearman coefficients, $r_s$. Also included are the stellar mass weighted values $\langle z_{\rm acc} \rangle$ and $\langle {\rm [Fe/H]} \rangle$ of the disrupted populations, the latter being computed at $z_{\rm acc}$ (see Section~\ref{ssec:av_prop}). For the MW we compute $\langle z_{\rm acc} \rangle$ and $\langle {\rm [Fe/H]} \rangle$ for the disrupted populations using data from \citet{Naidu2022}, both including I'itoi and Sagittarius (MW) and excluding them (MW$^{*}$). The parameters of the fit for the MW disrupted populations are from \citet{Naidu2022}, but we correct the offset to the same units used for the simulations. For the surviving populations in the MW we include the fit from \citet{Kirby2013}.} 
	\label{tab:MZR_fits}
\end{table*}

\begin{table*}
	\centering
	\begin{tabular}{|c|c|c|}
		\hline
		 & Disrupted & Surviving \\
		\hline
            
		\begin{tabular}{p{1cm}|p{1cm}|p{1cm}}
                Host & $ \langle z_{\rm{acc}} \rangle$ & $ \langle \rm{[Mg/Fe]} \rangle$\\
                \hline
                MW & 1.20 & 0.16 \\
                MW* & 1.80 & 0.20 \\
                G1 & 2.42 & 0.21 \\
                G23 & 1.76 & 0.13 \\
                G26 & 1.93  & 0.23 \\
                G29 & 2.11 & 0.16 \\
                G38 & 1.90 & 0.13 \\
                G42 & 1.98 & 0.17 \\
                \end{tabular} &
                \begin{tabular}{|p{1cm}|p{1cm}|p{1cm}|}
                Slope & Offset & $r_s$ \\
                \hline
                $-0.03$ & $~~0.41$ & -0.33 \\
                $~~0.02$ & $~~0.05$ & -0.09 \\
                $-0.13$ & $~~1.30$ & 0.75 \\
                $-0.08$ & $~~0.98$ & -0.52 \\
                $~~0.06$ & $-0.30$ & 0.54 \\
                $-0.06$ & $~~0.85$ & 0.60 \\
                $-0.13$ & $~~1.30$ & -0.70 \\
                $-0.11$ & $~~1.11$ & -0.86 \\
                \end{tabular}  & 
                \begin{tabular}{|p{1cm}|p{1cm}|p{1cm}|}
                Slope & Offset & $r_s$ \\
                \hline
                $-0.06$ & $0.50$ & -0.54 \\
                $-0.06$ & $0.50$ & -0.54 \\
                $-0.07$ & $0.51$ & -0.29 \\
                $-0.02$ & $~~0.12$ & -0.21 \\
                $-0.10$ & $~~0.65$ & -0.32 \\
                $-0.01$ & $-0.01$ & -0.70 \\
                $-0.01$ & $~~0.13$ & -0.30 \\
                $-0.02$ & $~~0.14$ & -0.04 \\
                \end{tabular}  \\
        
		\hline
        
	\end{tabular}
        \caption{Similar to Table~\ref{tab:MZR_fits}, but for the fits to the [Mg/Fe] -- ${M}_{*}$ relations corresponding to the six hosts in the $s_6$* subsample and to the observations. The $\langle {\rm [Mg/Fe]} \rangle$ values for the disrupted populations in the MW are computed using the data from \citet{Naidu2022}. The slopes and offsets for the relations of the disrupted and surviving populations in the MW are also computed using data from \citet{Naidu2022}.}
	\label{tab:alpha_vs_Mstar_fits}
\end{table*}


\bsp	
\label{lastpage}
\end{document}